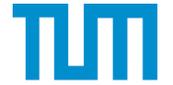

# TUM

TECHNISCHE UNIVERSITÄT MÜNCHEN
INSTITUT FÜR INFORMATIK

## Model-based Hazard and Impact Analysis

Sonila Dobi, Mario Gleirscher, Dr. Maria Spichkova, Prof. Dr. Peter Struss

TUM-I1333

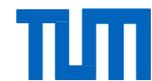

**Model-based Hazard and Impact Analysis**
(Modellgestützte Gefährdungsanalyse und Risikobewertung)

Sonila Dobi
Mario Gleirscher
Dr. Maria Spichkova
Prof. Dr. Peter Struss

Technische Universität München
Institut für Informatik
Chair of Prof. Dr. Dr. h.c. Manfred Broy
Software & Systems Engineering

# Abstract

*Hazard and impact analysis* is an indispensable task during the specification and development of safety-critical technical systems, and particularly of their software-intensive control parts. There is a lack of methods supporting an effective (reusable, automated) and integrated (cross-disciplinary) way to carry out such analyses.

This report was motivated by an industrial project whose goal was to survey and propose methods and models for documentation and analysis of a system and its environment to support hazard and impact analysis as an important task of safety engineering and system development. We present and investigate *three perspectives* of how to properly

- encode safety-relevant domain knowledge for better reuse and automation,
- identify and assess all relevant hazards, as well as
- pre-process this information and make it easily accessible for reuse in other safety and systems engineering activities and, moreover, in similar engineering projects.

The *first perspective* focuses on the transition from informal to a formal, model-based representation of knowledge about hazards and system requirements.

The *second perspective* provides a methodology to identify and treat hazards based on a state-machine model of the considered system.

The *third perspective* shows a tool-supported procedure for modeling faulty behaviors of both, physical and software components in a qualitative way and for automatically determining their impact based on the structural description of the physical and computational/software parts of the system and a model of the environment.

All perspectives are shown in their characteristics and capabilities by means of a case study on a *drive train in the commercial road vehicle* domain.

**Acknowledgements** We sincerely thank our project partners at ITK Engineering AG for many informative discussions and an effective collaboration throughout the project, Dr. Alessandro Fraracci for his contributions to the project, and OCC'M Software GmbH for providing Raz'r [OCC95] as the software basis for the automated safety analysis solution. We also thank Klaus Lochmann for a distinguished review including many helpful remarks on the manscript. Finally, sincere gratitude goes to our student assistants for converting important source documents.

# Contents

# 1 Motivation and Overview

This section gives an overview of current challenges of the safety domain and a previously performed collaboration with industry coining the central motivation for this report.

## 1.1 Challenges in System Safety Analysis

Cyber-physical systems are widespread in safety-critical domains such as, e. g., vehicles, production machinery, aircraft or medical devices. Failures of these systems may lead to considerable loss of money or even endanger human lives. This emphasizes the importance of correct behavior of these systems, which can be improved through analysis techniques, mostly in terms of formal verification [Cam10]. Formal analysis is facilitated by a formal model. A suitable modeling theory for these systems helps in their development, maintenance, simulation, and verification.

At a high level, the model of a cyber-physical system may not explicitly distinguish whether its subsystems or components are software or physical components, and they may be represented in a uniform way, e. g., as black boxes with a mapping from inputs to outputs or as transition systems. Often, these models try to capture the intended function of a system, rather than its entire possible behavior. For instance, in early phases of design, it may not yet have been decided whether a certain sub-system will be realized by software, a physical system, or a combination of both. Software components operate in discrete program steps, while the physical components evolve over time intervals following physical constraints. From our experience in a number of industrial collaborations [SC12], we know that system descriptions should not get too complicated and hardly readable. Hence, when modeling cyber-physical systems, one challenge is to find a proper abstraction of physical and computational phenomena.

However, when the behavior has to be analyzed in detail, the different nature of software and physical components will often require models that appropriately capture the physical phenomena that determine system behavior. This is even mandatory when the consideration of faulty behavior is involved, as e. g., in diagnosis, testing, or safety analysis: The variety of faults in software components can mostly be reduced to erroneous (manual or automated) transformations on the path from requirements to the executables (including inappropriate requirements). Beyond that, faults of physical components are much more subject to quite complex physical phenomena, such as e.g., wearout or unforeseen types of stress. However, this still makes it possible to enumerate and model at least the respective fault classes. As opposed to software components, most physical systems, such as components in electrical, mechanical, hydraulic, and pneumatic circuits, cannot be appropriately modeled by simple input-output behaviors. Even if they have an intended preferred cause-effect direction under nominal system behavior, this may be perturbed under the presence of a fault.

Having such characteristics in mind, a safety engineer has to gain understanding of hazards and perform safety-oriented validation of the specification. This includes understanding the relationship between software, electronic hardware and physical components' faults, and their consequences in terms of hazardous system failures and, finally, their impacts on the system environment. But the relationship between system safety goals and component safety and reliability requirements is often not formally captured. This makes it difficult to relate software, electronic or physical component misbehavior to hazards at the system level [Lev12].

Another issue is that technical systems are mainly modeled as glass-boxes to tie defects and their propagation to their component structure and to support detailed design for reliability. In other words, single phases of system operation (i. e., single transitions, linear event causal chain) are regarded. This is important but not enough as it detracts from the investigation of system behavior before and after critical events and the consideration of temporally distant or even external causal factors aside from system deficiencies. In other words, multiple phases of system operation (i. e., multiple transitions, non-linear causal chain) are regarded.

Hazard knowledge has hardly ever been systematically transferred to interdisciplinary, qualitative behavioral system models describing interaction between the system and its environment. At our chair, we work on approaches [Bro12, SF11] based on system models that are suitable to be *reused* and *checked for correctness, completeness and consistency* w.r.t. the tasks of safety engineering. Section 1.2 provides a short overview of related work that also aims to meet these challenges.

## 1.2 Overview of State-of-the-Art Safety Analysis

Figure 1.1 depicts approaches to *hazard analysis* according to their direction towards effects or causal factors of hazards. A more detailed version of the following overview with complete citations, references and remarks has been published in [Gle13].

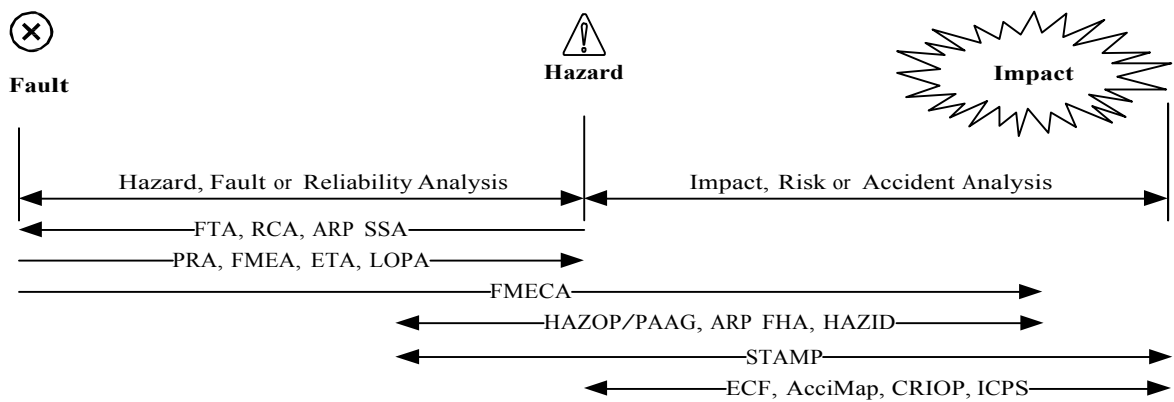

Figure 1.1: Approaches to hazard analysis for system safety assessment

**Defect Classification and Representation** Because of the variety in their perception, defects are categorized and modeled along technology and task specific, non-standard criteria, e.g., to compare system testing approaches, for fault-tolerance or reliability analysis, for mutation or fault-injection techniques [SF11].

**Top-Down from Hazard to Fault** This direction can be carried out using deductive techniques like, e.g., static or dynamic fault tree (FTA) [DBB92] or root cause (RCA) analysis. Some of the defect models are based on the data-flow architecture of a system, others apply state machine models extended with fault variables and ports for each system component.

**Bottom-Up from Fault to Hazard** This direction can be addressed by inductive methods like, e.g., failure mode, effect (and criticality) analysis (FME(C)A), event tree analysis (ETA) or, similarly, layer of protection analysis (LOPA). Hazard and operability analysis (HAZOP or PAAG) takes particular account of controllability by humans. There are elaborate approaches like probabilistic risk assessment (PRA) or informal, early-stage methods like hazard identification (HAZID) or preliminary hazard lists.

Model-based systems technology from Artificial Intelligence has been applied to automate the generation of effects (i. e., hazards) for FMEA from a model of the (faulty) system. The AutoSteve system [Pri00] was specialized on performing FMEA of electrical car subsystems and has been transferred into a commercial product (Capital SimCertify, [Gra]). The AUTAS project developed a generic FMEA tool with applications to electrical, hydraulic, pneumatic, and mechanical systems in aeronautic systems [PCB$^+$04] and has also been applied to automotive subsystems [SF12a].

**Between Hazard and Impact** To understand hazards and their risks in terms of impacts, methods like, e.g., events and causal factors (ECF) [BC95] or AcciMap [SR02] consider causal factors of all, environment, user and system. This includes system operations or use cases (e.g., driving missions and situations), operational incidents or damage scenarios (e.g., car accidents) as well as the physical system interface. Crisis intervention in offshore production (CRIOP) [JBS$^+$11] assesses the interface between human operators and technical systems within offshore control rooms to uncover obstacles for accident response. The system-theoretic accident model and processes (STAMP) [Lev12] perceives safety as a control problem in a socio-technical system, i.e., a collaboration of humans and technical systems. STAMP classifies human errors, identifies *inadequate control* beyond system failures and derives required constraints.

**Safety Engineering Guidance and Standards** Hazard analysis is a vital early step recommended by general or domain-specific safety standards, e.g., IEC 65108 for general mechatronics, ISO 26262 for automotive control, EN 50128 for train control or DoD MIL-STD-882D for technical systems. They consider the whole safety process including quantitative risk assessment for the regarded kinds of systems to avoid unwanted relationships between system safety goals and subsystem or component requirements. IEC 65108 concerns of preliminary hazard and risk analysis (PHA). The SAE aircraft recommended practices (ARP) 4754 and 4761 advise the steps of functional hazard assessment (FHA) based on a function list regarding failures and crew actions, followed by preliminary and final system safety/reliability assessment (SSA), which among FTA requires a mixture of techniques.

## 1.3 Project Goals

This report was inspired by the project *Efficient Hazard and Impact Analysis for Automotive Mechatronics Systems* supported by our industrial partners from *ITK Engineering AG*[1], Stuttgart-Vaihingen. The core activities of this project were carried through between May and December 2012. The goal of the project was to survey and propose methods and models for concise and reusable documentation of a system and its operational context (environment) in order to support system development, in particular, hazard and impact analysis—comparable to hazard analysis and risk assessment (HARA) in ISO 26262 part 3—as an important task of safety engineering. The proposed models and methods should capture enough details to fit their purpose and be restricted and structured in order to support efficient reuse and automation. In order to aim at the development of an appropriate safety analysis methodology, any proposed approach should comply as much as possible with the following criteria:

- Comprehensibility of models by educated control, safety and reliability engineers,
- Reasonable effort for model creation, use and reuse (maintenance or transformation),
- Manageable complexity by modular model decomposition,
- Ratability of relative completeness of specific analysis steps,
- Ratability of coverage for the verification of safety functions,
- Smooth transition into engineering practice by a multi-step and multi-view methodology.

Primary input to the project was an FMEA documentation (denoted as *provided data*, see Section 3) representing preliminary results of the hazard and impact analysis of a commercial road vehicle drive train (Sections 2 and 4.3). In addition to this input, the work at hand could gain from experiences and knowledge from other projects as well as efforts at the chair in the automotive and embedded software domain. In the mentioned case study, we explore three approaches introduced in Section 1.4, each taking a different perspective and focussing different sub-tasks of model-based hazard and impact analysis.

## 1.4 Overview of the Report

In this report, we focus on software and systems engineering aspects as well as on qualitative physical modeling and fault propagation. Following the project goals described before, we discuss three perspectives (Figure 1.2) in model-based, reusable safety (i. e., hazard and impact) analysis:

---
[1] www.itk-engineering.de

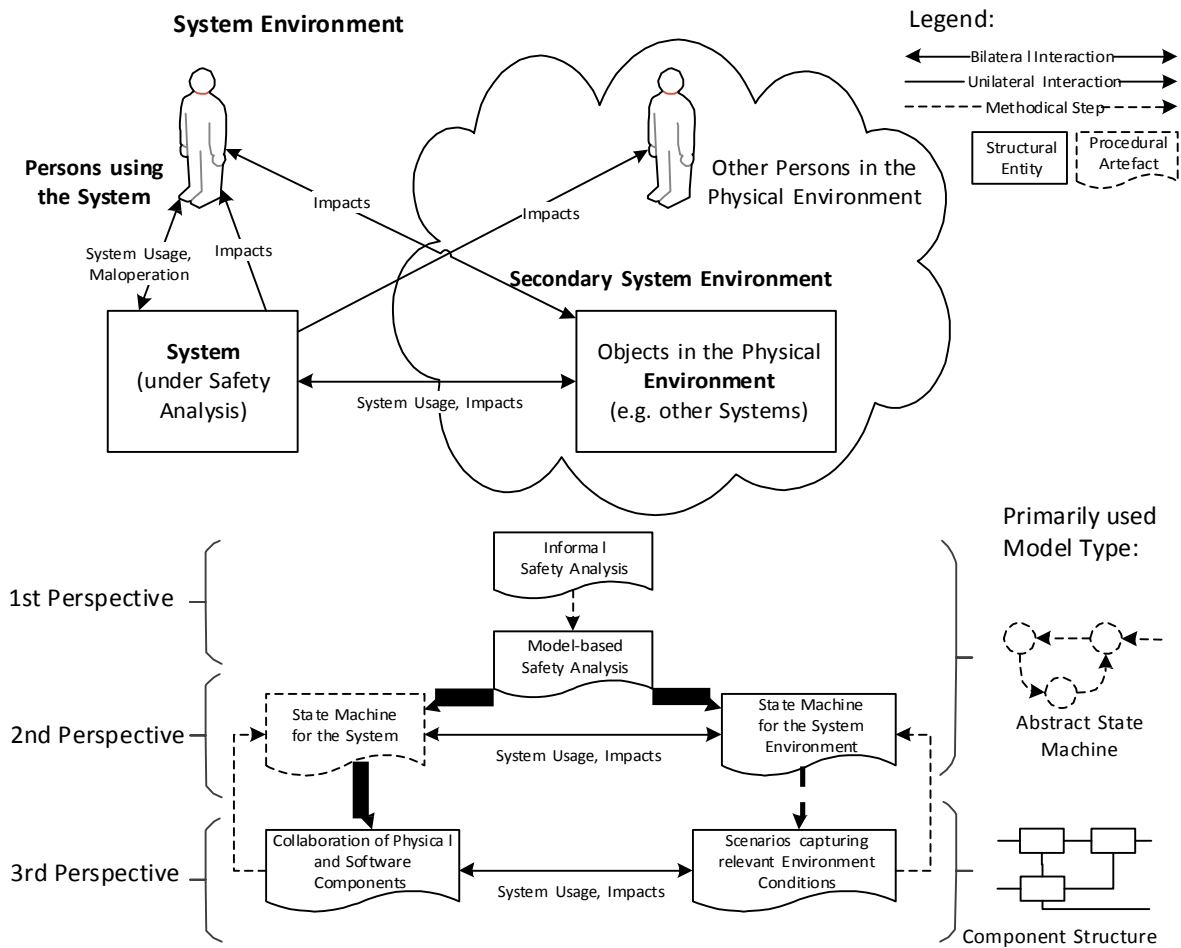

Figure 1.2: Overview and relationships between the three perspectives

**The first perspective** regards the *formalization of behavioral system properties* initially given in an informal manner. These properties can be *positive or wanted* (i. e., requirements) as well as *negative or unwanted* (i. e., hazards)—both are dual in many cases. This method is *abstract* in the sense that it does not care about the kind of properties—properties of computational or physical processes. The main ideas were presented in [SHT12]. Here, a new aspect of this representation is introduced: how to use it for the specification of hazards. Further details are described in Chapter 3.

**The second perspective** regards *safety analysis from a functional view* of both, the system and its environment. Hence, it focuses on system functionality, interface behavior and interaction between system and environment using state machine modeling. This method is *less abstract* than the previous one but neither splits system properties into virtual and physical ones, the model implicitly contains information about potential behavioral differences. For this, it deals with state machine models representing faulty behavior. The state machine model is the result of manual analysis and specific to the functionality of the system under consideration. Further details are described in Chapter 4.

**The third perspective** approaches hazard analysis by automatically deducing the presence or absence of hazards based on behavior models of system components and the component structure. Also the impact of hazards on the environment is generated automatically based on an abstract model of the environment. The approach builds on previous work on automated FMEA of physical systems [Str08], uses qualitative relational behavior modeling, and extends them to include a small set of high-level (fault) models of the software components. In contrast to the other two perspectives, the component structure is represented explicitly and used to compose the whole system model automatically from a component library. Based on this, also the analyses of hazards and impacts are computed automatically. Further details are described in Chapter 5.

**Cross-cutting Aspects** Common to all three perspectives is the *interpretation* of the provided data by means of formally founded system models: The first perspective (Section 3) gives explicit guidance for this. The second and third perspectives provide a modeling framework (Sections 4.1 and 5.2) where this has to be done manually. Along with this goes the *identification of a proper system boundary* (Section 4, Table 3.3 and Figures 5.1 and 5.2) and structure (Section 5.5.3). This is the basis for capturing *relationships* between hazards and system functions (Section 4) as well as the causal chains from faults and hazards up to their impacts (Section 5). Finally, the perspectives aim at *enhancing* the provided data by additional models suited for reuse and automation (Sections 4 and 5). Beginning with Section 2, we discuss a case study on a *Commercial Road Vehicle* from these three perspectives.

# 2 Introduction to the Case Study: Commercial Road Vehicle Safety

Our industrial project partner ITK Engineering AG [ITK12] proposed the domain of commercial road vehicles (*trucks* for short) as the subject of our case study, particularly, *drive trains* of trucks. The structure of a drive train is sketched in Figure 2.1.

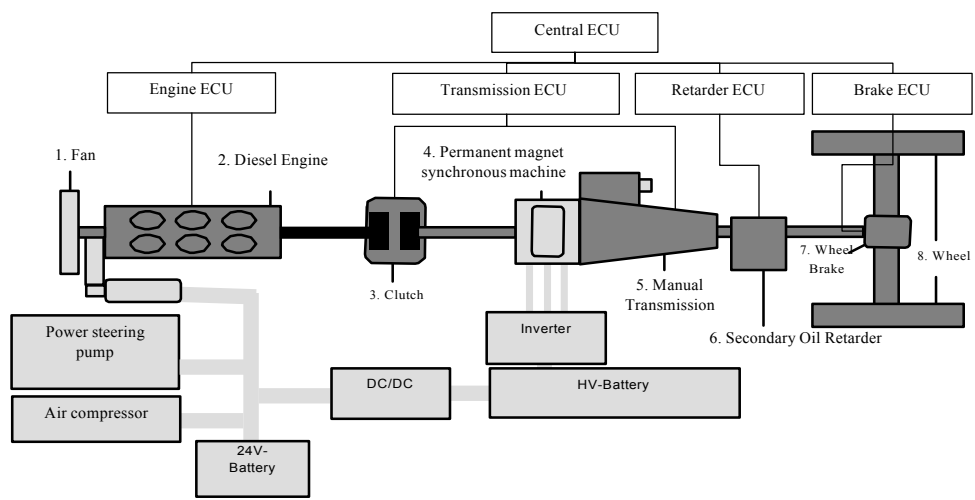

Figure 2.1: Physical and control component structure for the truck drive train

The main subsystem of a drive train (in dark gray color) comprises

- the diesel *engine* (2), equipped with a fan (1), providing an accelerating or braking torque,
- the *clutch* (3) which may interrupt the propagation of torque,
- the *transmission* (5) and the *generator* (4) allowing to switch between forward and reverse torque (and idling),
- the *retarder* (6), a braking device that, when applied, counteracts the rotational motion through a propeller moving in oil,
- the drive *axle* together with the wheel (8), which transforms rotational acceleration into translational acceleration (and vice versa), and
- the *wheel brakes* (7).

Components are controlled by specialized *electronic control units* (ECU), which communicate with a central ECU that processes, for instance, the driver demands. The light-gray components are related to electrical aspects and have not been treated in the first phase of the project. The industrial partner also supplied us with exemplary problems and results in safety analysis. Table 2.1 and Table 2.2 show part of the lists of hazards and the relevant environment conditions to be analyzed (provided by ITK Engineering AG [ITK12]), respectively. Part of the work in the case study was to turn these informal descriptions into a semi-formal representation that would support automated model-based reasoning (see Section 3).

| ID | Hazard |
|---|---|
| 1 | Unintended start |
| 2 | Start opposite to the intended direction of travel |
| 3 | Unintended acceleration |
| 4 | Unintended deceleration |
| 5 | Loss of steering power |
| 6 | Failure of the braking system |
| 7 | Electric shock |
| 8 | Fire |

Table 2.1: List of hazards

| ID | List of relevant driving situations |
|---|---|
| 1 | Car wash |
| 2 | Refueling (petrol/diesel) |
| 3 | Parking on a slope |
| 4 | Crossing an intersection |
| 5 | Passing cars parked on the roadside |
| 6 | Dense traffic with vulnerable persons on the road (e. g., pedestrians, cyclists) |
| 7 | Deceleration at, e. g., traffic lights, stop sign |
| 8 | Driving in reverse |
| 9 | Driving in roundabouts |
| 10 | Freeway exit, deceleration |
| 11 | Approaching a traffic jam |

Table 2.2: List of relevant driving situations as a part of the environment conditions

Other tables summarize results of the manually generated analysis (provided by ITK Engineering AG [ITK12]). In Table 2.3, each row links a hazard (column 2), which are characterized in column 3, with the components whose faults may be responsible for it (column 4). Table 2.4 captures the results of the impact analysis: for a hazard (column 2) and particular environment conditions (column 3), the impacts, e. g., persons injured by the vehicle, are determined (column 6).

| ID | Hazard | Additional Information | Components with Potential to trigger the Hazard (additional conditions in parentheses) | Physical System Function |
|---|---|---|---|---|
| 1 | Unintended start | Vehicle standing, unintended torque on drive axle (independent of direction) | **Generator** (gear engaged), **Clutch** (gear engaged, engine is running or starter is actuated) | Drive |
| 2 | Unintended start-up less than 5 km/h | Vehicle standing, unintended torque on drive axle (independent of direction) | **Starter** (clutch engaged) | Drive |
| 3 | Start opposite to intended direction | Vehicle standing, torque on drive axle opposite to intended direction | **Generator**, **Clutch** (clutch open) | Drive |
| 4 | No start on demand | Vehicle standing, missing torque on drive axle | **Generator** (gear engaged, clutch open), **Clutch** (gear engaged, engine running) | Drive |
| 5 | Unintended acceleration | Vehicle standing, unintended torque on drive axle in direction of wheel rotation | **Generator** (gear engaged), **Engine** (clutch and gear engaged) | Accelerate |
| 6 | No acceleration on demand | Vehicle moving, no torque on drive axle in direction of wheel rotation | **Retarder**, **Generator**, **Clutch** (engine running) | Accelerate |
| 7 | Unintended deceleration | Vehicle moving, unintended torque on drive axle opposite to wheel rotation | **Retarder**, **Generator**, **Engine** (clutch engaged), **Clutch** | Accelerate |
| 8 | No deceleration on demand | Vehicle moving, no torque opposite to wheel rotation on drive axle | **Retarder**, **Generator**, **Engine** (clutch engaged), **Clutch** (engine braking, open clutch results in loss of engine brake) | Brake |
| 9 | Unintended stopping | Vehicle moving, unintended torque on drive axle opposite to wheel rotation | **Generator** (gear engaged), **Engine** (clutch closed, engine off (it brakes)) | Stop |
| 10 | Delayed stopping (operational brakes) | Vehicle decelerates until stop, unintended torque on drive axle in direction of wheel rotation | **Generator** (gear engaged), **Clutch** (engine running) | Stop |
| 11 | Unintended engine start | | **Generator** (clutch engaged, gear in neutral position), **Starter** | Start engine |
| 12 | No engine start on demand → vehicle operation not possible | | **Generator** (clutch engaged, gear in neutral position), **Starter** (clutch open), **Engine** (no injection), **Clutch** (neutral gear) | Start engine |

continued from previous page

| ID | Hazard | Additional Information | Components with Potential to trigger the Hazard (additional conditions in parentheses) | Physical System Function |
|---|---|---|---|---|
| 13 | Unintended loss of 24 V supply → No supply to power steering pump | | **24 V battery**, **DC/DC converter**, **HV battery** | 24 V power supply for auxiliary equipment |
| 14 | Gear change not possible | | **Generator**, **Engine**, **Clutch** | Support of gear change |
| 15 | Electric shock | | **HV-Battery** **Inverter** **DC-DC-Converter** **Generator** | Driving, parking, ... |

Table 2.3: Results of manual hazard analysis (translated from external project documentation in German [ITK12]): hazards, description of hazards, faulty components and violated system functions

| ID | Hazard | Environment condition | Impact | E | S | C | ASIL |
|---|---|---|---|---|---|---|---|
| H1.1 | Unintended engine start | Vehicle is in pit lane | Injury of persons in pit lane | E2 | S2 | C3 | B |
| H1.2 | Unintended engine start | Workshop | Injury of mechanics | E3 | S2 | C3 | QM |
| H1.3 | Unintended engine start | Vehicle in starting grid | Collision with other racing cars | E2 | S1 | C2 | QM |
| H1.4 | Unintended engine start | Accident on race course | Injury of persons (e. g., racing teams) | E2 | S2 | C2 | QM |
| H2.1 | No engine start on demand | Pit lane engine stalling | Subsequent cars colliding | E2 | S1 | C1 | QM |
| H2.2 | No engine start on demand | Vehicle in starting grid, engine stalls just before the start | Subsequent cars colliding | E2 | S1 | C2 | QM |

Table 2.4: Results of manual impact analysis for racing car scenarios (translated from external project documentation in German [ITK12]): hazards occurring in a certain environment condition cause a specific impact. They are assessed with respect to likelihood (E), severity (S), controllability (C) and classified according to ISO 26262 using the Automotive Safety Integrity Level (ASIL) scheme. Basic hazards are listed in Table 2.1

# 3 Formalizing Hazards as Behavioral Properties

Section 3.1 provides an overview of the concepts used and the procedure followed in this perspective. Sections 3.2 and 3.3 exemplify this procedure in the context of our case study.

## 3.1 Overview of the Concepts and the Procedure

As an input for our case study, we take the *hazard table* among the *provided data* [ITK12] for a commercial road vehicle (Sections 1.3 and 2). The purpose and contribution of this perspective is a natural language-oriented interpretation of the informal *hazard table* in order to optimize it and to find out whether some hazards are only special cases of other ones. The focus lies on accurate transcription rather than optimal modeling.

We can start with an *informal specification* of behavioral properties or with a general model of the *system under consideration*. Then we specify the system behavior in a *semi-formal* way by transforming the natural-language requirements either into a structured form using pre-defined *textual patterns* as presented in [Fle08], or into *message sequence chart* representation (MSC, [HT03]) according to the approach presented in [Spi10]. The purpose of using MSCs for system specification is to obtain an overview in comparison to simple textual patterns. However, it better suited for systems with a need to discuss many complex interaction scenarios being part of major system use cases. In the presented case study, we restrict our approach to using textual patterns as described below.

Subsequently, a semi-formal specification can be translated into FOCUS [BS01], a formal framework for specification and development of distributed interactive discrete-event systems, a general class of systems capturing many aspects of cyber-physical systems. A formal representation enables the automated verification of system architecture and realization against system requirements by translation of both artifacts into the language of a theorem prover, e. g., Isabelle/HOL [NPW02] via the framework "FOCUS on Isabelle" [Spi07].

An informal specification consists of a set of words, which can be distinguished into two categories: content words and keywords (relation words). Content words are system-specific words or phrases, e. g., *"system is initialized"* or *"off-button is pressed"*. The set of all content words forms the system interface, which can be understood as a domain specific, system-dependent vocabulary that has to be defined in addition. Keywords are domain-independent and form relationships between content words such as e. g., "if", "then" and "else". A semi-formal specification consists of a number of requirements described using the following tex-

| Mode Name | Explanation |
|---|---|
| Off15 | The clamp 15 of the vehicle is off. |
| On15 | The clamp 15 of the vehicle is on. |
| Stop | The vehicle does not move, clamp 15 of the vehicle is on. |
| Drive | The vehicle is driving, clamp 15 of the vehicle is on. |
| Acceleration | The vehicle accelerates, clamp 15 of the vehicle is on. |
| Deceleration | The vehicle decelerates, clamp 15 of the vehicle is on. |
| ConstDrive | The vehicle is driving at const. speed, clamp 15 of the vehicle is on. |

Table 3.1: Explanation of the vehicle mode names

tual pattern, which can be better understood by engineers unfamiliar with formal methods:

> *WHILE*   Some state or system mode
> *IF*      Some event occurs or some state changes
> *THEN*    Some event occurs or some state changes
> *ELSE*    Some event occurs or some state changes

An event describes a *point in time*, in which the system observes or does something; the duration of the event is not important, e. g., "driver presses a button". A state describes a system or component state *within some time period*, e. g., "a button is pressed". Using such a description to structure the information from an informal specification, we can find out missing information at lower cost. Furthermore, we identify possible synonyms that must be unified before proceeding to a formal specification. Analysis of the semi-formal specification should also yield sentences, which need to be reformulated or extended.

## 3.2  Case Study: Modeling System Modes

First of all, we discuss specific system states considered as modes. Figure 3.1 present the hierarchy of them, where Figure 3.2 shows the corresponding general state transition diagram for a vehicle and serves, as a first step, for the modeling and analysis performed in the second perspective (Section 4). The meaning of the mode names is presented in Table 3.2.

Note, that the provided data (Section 2) only refers to the modes *Off15*, *Stop*, and *Drive*. This viewpoint lacks some more details.

## 3.3  Case Study: From Hazards to Safety Goals

The following list shows the transcriptions of the informal hazard list (Table 2.3) into semi-formal temporal or behavioral property assertions. We wanted to maintain traceability and verifiability within the original language, particularly, for our project partners. Hence, we included sentences and phrases from the original documents, all written in German, into the language analysis below. For readers not used to the German language, the translations of the corresponding parts of these documents are referenced to be looked up in Table 2.3.

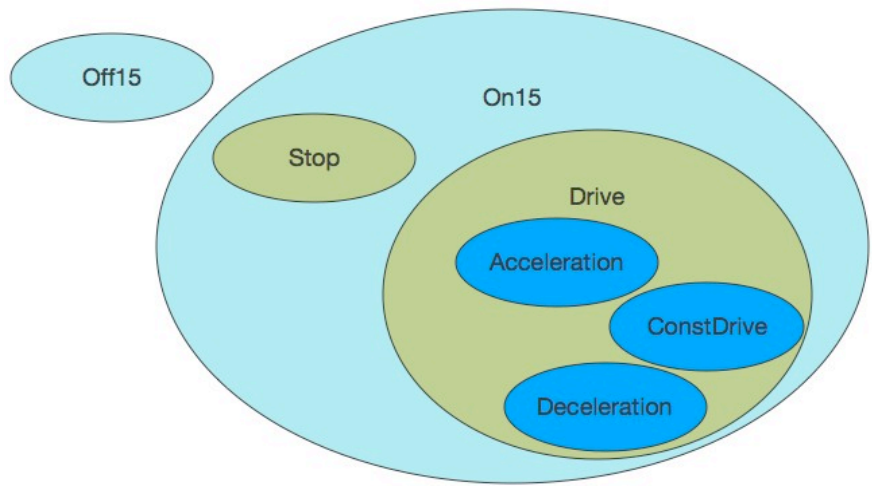

Figure 3.1: System Modes: Hierarchy

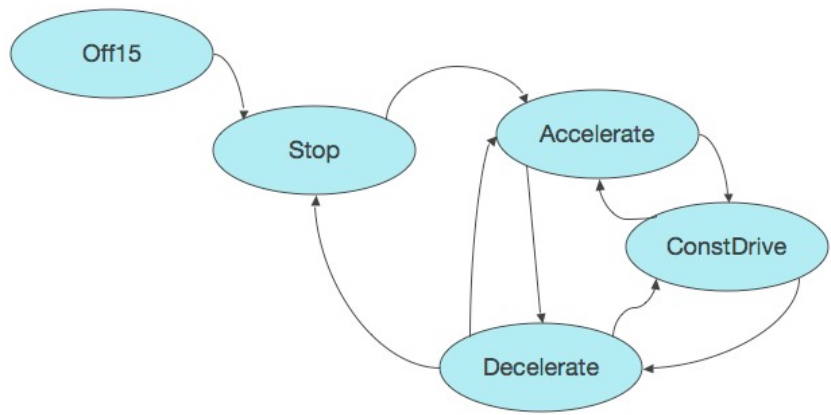

Figure 3.2: Mode transition diagram

**Hazard 1** (Table 2.3 ID1) *Ungewolltes Anfahren:*
  *Fahrzeug steht, ungewollte Momentenabgabe unabhängig von der Richtung an der Antriebsachse.*
  Reformulating this hazard to the corresponding safety property we obtain the following sentence:
  *Fahrzeug steht, ungewollte Momentenabgabe unabhängig von der Richtung an der Antriebsachse ist ausgeschlossen.*
  We can represent it semi-formally as follows:
  *IF Fahrzeug steht*

*THEN NOT (ungewollte Momentenabgabe)*

In English:
*IF the vehicle is standing*
*THEN NOT (unintentional torque output)*
More precisely:
(P1)
*IF the vehicle's mode is Stop AND NOT (torque output request)*
*THEN NOT (torque output)*
(Semi-)Formalization of hazard:
(H1)
*IF the vehicle's mode is Stop AND NOT (torque output request)*
*THEN (torque output)*

**Hazard 2** (Table 2.3 ID2) *Ungewolltes Anfahren bis 5 km/h:*
*Fahrzeug steht, ungewollte Momentenabgabe unabhängig von der Richtung an der Antriebsachse.*

The difference to H1 is a refined name and that we restrict the situation to the vehicle speed ≤ 5 km/h. Thus, we get the following hazard assertion:
(H2)
*IF the vehicle's mode is Stop AND NOT (torque output request)*
*THEN (torque output) AND (vehicle speed ≤ 5 km/h)*

The reason for this restriction was not fully clear. However, as property P1 also captures hazard H2, a restricted version of P1 is neither necessary nor acceptable.

**Hazard 3** (Table 2.3 ID3) *Anfahren entgegen der gewünschten Fahrtrichtung:*
*Fahrzeug steht, Momentenabgabe entgegen der gewünschten Richtung an der Antriebsachse.*
(Semi-)Formalization of hazard:
(H3)
*IF the vehicle's mode is Stop AND (torque output request)*
*AND (requested torque direction is R1)*
*THEN (torque output) AND (torque direction is opposite to R1)*

Note that the *requested direction* (see Figure 3.3) captures the direction of the vehicle. If *requested direction* has more values than *forward* and *backward* (e. g., *left*, *right*), we would have to explicitly define pairs of opposite values. However, we can specify the hazard *H*3 in more general:
(H3')
*IF the vehicle's mode is Stop AND (torque output request)*
*AND (requested torque direction is R1)*
*THEN (torque output) AND (torque direction /= R1)*
The corresponding property:
(P2)
*IF the vehicle's mode is Stop AND (torque output request)*
*AND (requested torque direction is R1)*
*THEN (torque output) AND (torque direction is R1)*

**Hazard 4** (Table 2.3 ID4) *Kein Anfahren auf Anforderung:*
*Fahrzeug steht, keine Momentenabgabe an der Antriebsachse auf Anforderung.*
(Semi-)Formalization of hazard:
(H4)
*IF the vehicle's mode is Stop AND (torque output request)*
*THEN NOT (torque output)*
The corresponding safety property is the generalization of the property *P2* (we do not care here of the vehicle direction).
(P2')
*IF the vehicle's mode is Stop AND (torque output request)*
*THEN (torque output)*

**Hazard 5** (Table 2.3 ID5) *Ungewollte Beschleunigung:*
*Fahrzeug fährt, ungewollte Momentenabgabe in Raddrehrichtung an der Antriebsachse.*
(Semi-)Formalization of hazard:
(H5)
*IF the vehicle is in the mode Drive AND NOT (torque output request)*
*AND (direction is R)*
*THEN (torque output) AND (torque direction is R)*
Please note that we cannot reformulate this hazard as follows:
(H5wrong)
*IF the vehicle is in the mode Drive AND NOT (torque output request)*
*AND (direction is R)*
*THEN (vehicle speed increases) AND (torque direction is R)*
because the vehicle speed can also increase because of the environment's influence, e.g. if the vehicle goes downhill. Moreover, if the vehicle goes uphill, its speed can stay constant or decrease even when hazard H5 occurs—essential to this hazard is exactly the problem of an unrequested torque output. However, we can reformulate[1] the hazard in the following way:
(H5')
*IF the vehicle is in the mode Drive AND NOT (acceleration request)*
*AND (direction is R)*
*THEN (torque output) AND (torque direction is R)*
We can generalize H5' omitting the information about the direction of vehicle (of vehicle's wheels):
(P3)
*IF the vehicle is in the mode Drive AND NOT (torque output request)*
*THEN NOT (torque output)*

**Hazard 6** (Table 2.3 ID6) *Keine Beschleunigung auf Anforderung:*
*Fahrzeug fährt, keine Momentenabgabe in Raddrehrichtung an der Antriebsachse auf Anforderung.* This hazard is dual to *H5*. Its (semi-)formalization:
(H6)
*IF the vehicle is in the mode Drive AND (torque output request)*
*AND (direction is R)*

---

[1] The Figures 3.3 and 3.4 show how the system boundary is defined to specify hazards and safety goals and which parts of this boundary are used in the various semi-formal assertions, e. g., H5 vs. H5'.

*THEN NOT ((torque output) AND (torque direction is R))*
This is equal to
(H6')
*IF the vehicle is in the mode Drive AND (torque output request)*
*AND (direction is R)*
*THEN NOT (torque output) OR (torque direction /= R)*
We can reformulate the hazard in the following way, analog to *H5'*:
(H6")
*IF the vehicle is in the mode Drive AND (acceleration request)*
*AND (direction is R)*
*THEN NOT ((torque output) AND (torque direction is R))*
The corresponding property is dual to the property *P3* and differs from the property *P2* only in the system mode. We also generalize it omitting the information about the direction of vehicle (of vehicle's wheels):
(P4)
*IF the vehicle is in the mode Drive AND (torque output request)*
*THEN (torque output)*

**Hazard 7** (Table 2.3 ID7) *Ungewollte Verzögerung:*
*Fahrzeug fährt, ungewollte Momentenabgabe entgegen der Raddrehrichtung an der Antriebsachse.*
(Semi-)Formalization of hazard:
(H7)
*IF the vehicle is in the mode Drive AND NOT (torque output request) AND (direction is R)*
*THEN (torque output) AND (torque direction is opposite to R)*
(P5)
*IF the vehicle is in the mode Drive AND NOT (torque output request) AND (direction is R)*
*THEN NOT (torque output) AND (torque direction is R)*

**Hazard 8** (Table 2.3 ID8) *Keine Verzögerung auf Anforderung:*
*Fahrzeug fährt, keine Momentenabgabe entgegen der Raddrehrichtung an der Antriebsachse auf Anforderung.*
(Semi-)Formalization of hazard:
(H8)
*IF the vehicle is in the mode Drive AND NOT (torque output request)*
*AND (directionReq is opposite to R) AND (direction is R)*
*THEN NOT ((torque output) AND (torque direction is opposite to R))*
The corresponding property:
(P6)
*IF the vehicle is in the mode Drive AND NOT (torque output request)*
*AND (directionReq is opposite to R) AND (direction is R)*
*THEN (torque output) AND (torque direction is opposite to R)*

**Hazard 9** (Table 2.3 ID9) *Ungewolltes Anhalten:*
*Fahrzeug fährt, ungewollte Momentenabgabe entgegen der Raddrehrichtung an der Antriebsachse bis zum Fahrzeugstillstand.*

This hazard can be seen as a special case (refinement) of the hazard $H7$, thus, if we have proven that the hazard $H7$ is excluded, we can omit the proving of $H9$, however, its proof is necessary only if it is to complicated to argue of $H7$. It can be represented in a (semi-)formal way as follows[2]:

(H9)
*IF the vehicle is in the mode Drive AND NOT (torque output request) AND (direction is R)*
*THEN (torque output) AND (torque direction is opposite to R)*
*AND (vehicle's mode is Stop)$^N$*

This hazard corresponds to the property $P5$ we define above.

**Hazard 10** (Table 2.3 ID10) *Verzögertes Anhalten (mittels Bremssystem):*
*Fahrzeug verzögert bis zum Fahrzeugstillstand, ungewollte Momentenabgabe in Raddrehrichtung an der Antriebsachse.*
The additional constraint *until the current speed is equal to 0* can be omitted here.

(H10)
*IF the vehicle's mode is Decelerate AND (direction is R)*
*AND NOT ((torque output request) AND (requested torque direction /= R))*
*THEN (torque output) AND (torque direction is R)*

This corresponds to the following property:

(P7)
*IF the vehicle's mode is Decelerate AND (direction = R)*
*AND NOT ((torque output request) AND (requested torque direction /= R))*
*THEN NOT(torque output) AND (torque direction is R))*

**Hazard 11** (Table 2.3 ID11) *Ungewollter Motorstart:*
*Klemme 15 aus, ungewollte Motorstart.*
The start of the motor corresponds in our model to the change of the system mode *Off15* to *On15*, more precisely, from mode *Off15* to *Stop*, therefore we can represent this hazard (semi-)formally as follows:

(H11)
*IF the vehicle is in the mode Off15 AND NOT Clamp15On*
*THEN the vehicle's mode is Stop*

The corresponding property:

(P8)
*IF the vehicle's mode is Off15 AND NOT Clamp15On*
*THEN the vehicle's mode is Off15*

**Hazard 12** (Table 2.3 ID12) *Kein Motorstart auf Anforderung:*
*Klemme 15 an, kein Motorstart auf Anforderung.*
This hazard is dual to the hazard $H11$:

(H12)
*IF the vehicle's mode is Off15 AND Clamp15On*
*THEN the vehicle's mode is Off15*

The corresponding property is dual to the property $P8$:

(P9)

---

[2] The notation $event^N$ denotes the proposition that *event* will take place in $N$ steps of the system, i. e., in $N$ time units.

*IF the vehicle's mode is Off15 AND Clamp15On*
*THEN the vehicle's mode is Stop*

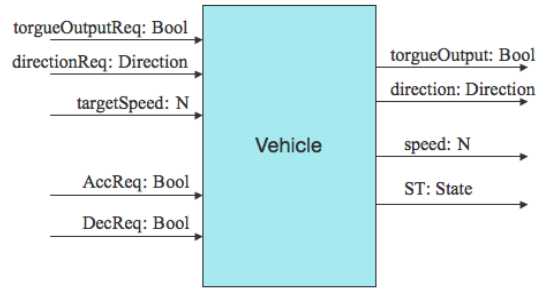

Figure 3.3: System interface (initial vocabulary)

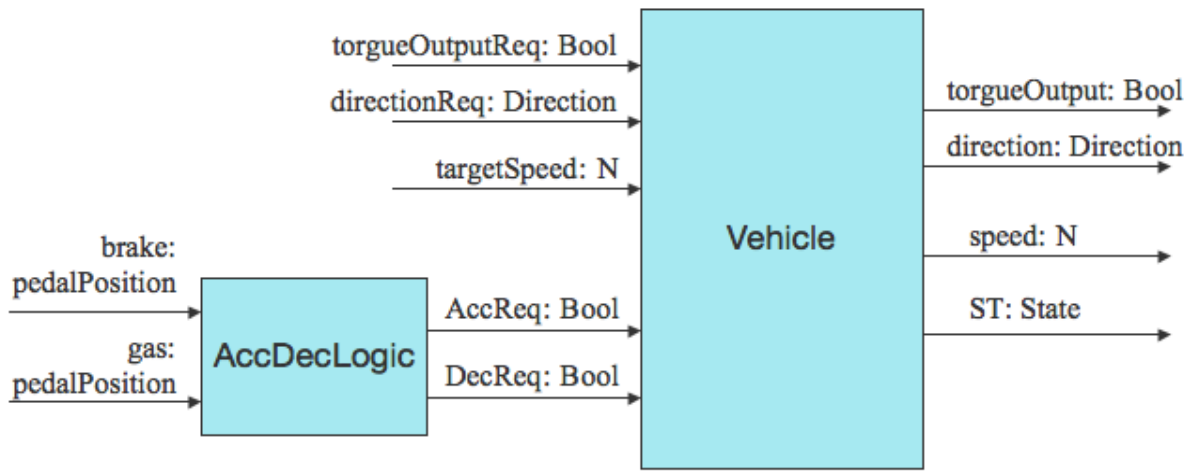

Figure 3.4: System interface (vocabulary after a refinement)

The transformation of the informal hazard assertions into semi-formal assertions using the textual patterns contributed in a more precise definition of the system boundary. Based on that, the Figures 3.3 and 3.4 represent the system interface in a graphical manner. For the interface ports, we define simple types:
$Direction = \{forward, backward\}$,
$Bool = \{true, false\}$,
$N = \{n \mid n \text{ is a natural number}\}$,
$State = \{off15, on15 = clamp15on, stop, drive, accelerate, constdrive, decelerate\}$,
$pedalPosition = \{released, pressed\}$.

3 Formalizing Hazards as Behavioral Properties23| ID | Formalized Hazard(s) | Abstracted Hazard | Corresponding Safety Property |
|---|---|---|---|
| 1 | $H1$ |  | $P1$ |
| 2 | $H2$ | $H1$ | $P1$ |
| 3 | $H3$ | $H3^t$ | $P2$ |
| 4 | $H4$ |  | $P2^t$ (generalization of $P2$) |
| 5 | $H5$ | $H5^t$ | $P3$ |
| 6 | $H6$ (dual to $H5$) or $H6^t$ | $H6^{tt}$ | $P4$ (dual to $P3$) |
| 7 | $H7$ |  | $P5$ |
| 8 | $H8$ |  | $P6$ |
| 9 | $H9$ | $H7$ | $P5$ |
| 10 | $H10$ |  | $P7$ |
| 11 | $H11$ |  | $P8$ |
| 12 | $H12$ (dual to $H11$) |  | $P9$ (dual to $P8$) |

Table 3.2: Relationships between the hazards semi-formalized based on Table 2.3 and corresponding safety properties

| **First Perspective** (Section 3) | **Comments** | **Second Perspective** (Section 4) |
|---|---|---|
| speed: N | speed of the vehicle | vehicle.speed |
| direction: Direction | direction of the vehicle | vehicle.direction |
| directionReq: Direction | requested direction of the vehicle | steeringwheel.pos |
| torqueOutput: B | current torque | forcemomentum.wheels |
| torqueOutputReq: B | requested torque | forcemomentum.extra |
| torqueDirReq: Direction | requested torque direction |  |
| torqueDir: Direction | current torque direction |  |
| brake: PedalPosition | used for modeling *accReq*, *decReq* | brakepedal.pos |
| gas: PedalPosition | used for modeling *accReq*, *decReq* | gaspedal.pos |
| targetSpeed: N | target speed of the vehicle | acc.targetspeed |
| Clamp15On: B | true corresponds to $On15$ | key.pos |
| - |  | vehicle.pos |
| - |  | vehicle.load |
| - |  | keyslot.status |
| - |  | clutchpedal.pos |
| - |  | gearlevel.pos |
| - |  | brakepedal.force |
| - |  | stopbrake.status |

Table 3.3: Interface ports: Vocabulary bridging the gap to the second perspective (Section 4)

## 3.4 Summary

The hazard list of Table 2.3, which is based on Table 2.1, has been processed as follows: Out of the 15 listed hazards, 11 have been transformed using textual patterns. Hazard H9 has been added as a refinement of H7. To reduce redundancy, due to lack of knowledge or to keep the model simple enough for a case study, the hazards "no acceleration on demand", "involuntary loss of 24V", "gear change not possible" and "electric shock" have been left out. This does not harm the validity of the model and the method. The result of analyzing the above 12 hazards is presented in Table 3.2. Although we maintained original language assertions for traceability reasons, the presented approach itself can be applied to assertions in English or other natural languages.

In this perspective, we showed how to transform informal text into controlled language using textual patterns. This improved our understanding of the system interface including global system states and modes, a more precise definition of interface ports as well as a more complete definition of wanted or unwanted behaviors observable over these ports. To switch to the second perspective (Section 4), we can orientate on this interface definition as well as the list of hazards and safety goals represented in a semi-formal way. Figure 3.3 shows the resulting specification of the system interface gained during this transformation. Table 3.3 relates the first and the second perspective (Section 4) by referring to the interface ports.

# 4 Hazard Analysis based on State Machines

The second perspective shows a *method* consisting of a *framework* and a *procedure* for qualitative behavioral modeling and hazard analysis. This perspective is based on a variant of state machines and exemplified by a commercial road vehicle as introduced in Section 2. The method was developed in the course of the project mentioned in Section 1.3 and has already been published in isolation and in more detail in [Gle13]. We show an extract of this article, summarize the most important concepts of the method and exemplify it by embedding it into the current case study.

## 4.1 Framework Concepts

To address the problems and challenges stated in Section 1, the *framework* comprises a *qualitative behavioral model* of the system's functionality, its environment and a set of *property assertions*, altogether used as a specification. We denote the functionality of the environment by $f^E$ and the functionality of the system under consideration by $f^I$. In the following, we use $* \in \{I, E\}$ and $f^*_{label}$ will denote a part of these functionalities identified by label. The behavioral model is based on a system interface defined on the basis of variables also called *interface phenomena* and consists of three *functional aspects*:

- *Usage functionality* $f^*_{use}$ (also called *nominal* functionality),
- *defective functionality* $f^*_{fail}$ (for explicit defect modeling, e.g., to describe failures), and
- *safety functionality* $f^*_{save}$ as an enhancement thereof to avoid or mitigate hazards.

At the system side, parts of the functionality are called *usage functions*, at the environment side *environment tactics*. The portion of a function or tactic belonging one of the aspects is called *functional fragment*. To characterize important points in time, some valuations of the interface phenomena are used as local situations in property assertions and simulations of the model. Based on this specification, a hazard combines a hazardous element with an initiating mechanism to threaten a target. The risk consists in a potential negative outcome from this mechanism's performance, i.e., an impact on the target or, synonymously, a mishap for it. Risks are usually quantified by severity and probability values.

## 4.2 Procedural Steps

Based on Figure 4.1, the *procedure* is explained in the following:

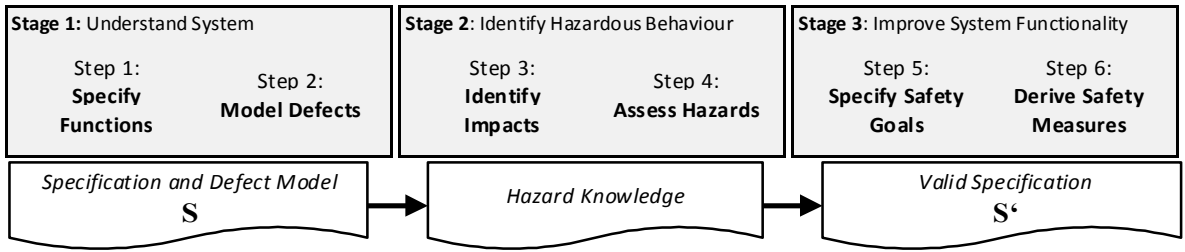

Figure 4.1: A procedure in six steps for using the framework (adopted from [Gle13])

**Step 1** *Specify the Usage Functionality* — Guiding Questions: *What is considered as nominal behavior?* Transform the informal function definitions known from requirements engineering into *use cases* and the set of informal property assertions into semi-formal (e.g., as described in Section 3) or formal assertions. Determine the outmost system boundary and its interface phenomena (right column of Table 3.3). Transform the use cases into a *hierarchy of usage functions* and identify *modes of operation* and *operations* to derive a model of *interaction* via *usage functions* and *environment tactics*. For each leaf of the hierarchy, identify *actions* (transitions), their repetition and order by the help of *modes*.

Methodical details on how use cases can be elaborated are described in, e.g., [Coc00]. Further information on the specification of a function hierarchy can be taken from [Bro05, Rit08, Bro10, HM03].

**Step 2** *Model the Defective Functionality* — Guiding Questions: *Which failures are possible? Which actions could be performed in an unexpected manner?* Apply several strategies to derive a defect model: introduction of indeterminacy or physical side effects, fault knowledge and other mutations. This step avoids the implicit assumption that hazards are always failures, as it was made in the first perspective (Section 3). Hence, we separate hazard analysis from reliability analysis to a certain extent.

**Step 3** *Identify Potential Impacts* — Guiding Questions: *Which impacts are possible? Which actions could be performed in an impacting manner?* Determine *physically relevant* operations and identify *conditions of harming events* based on interface phenomena, e.g., areas which could get *contaminated* or where objects *collide*, get *sounded*, *glared* or *shot* ; places where objects could get *clamped*, *sheared*, *scraped* or *cut* ; surfaces where objects could get *burned*, *vibrated*, *electrically shocked* or *dissolved*. Derive *candidate hazards* which indicate *hazardous executions* or *suppressions* and separate *hazardous* from *safe performance* of the regarded operation. Define local situations.

*Proposed Formal Instrumentation:* We express and specify impacts as constraints on the available interface phenomena or state variables known from Step 1. A constraint can be any first-order predicate calculus formula using terms (incl. arithmetic) and inequalities (incl. equalities). In order to express and specify hazards, we propose to use *linear temporal logic*. This will be exemplified in the case study in Section 4.3 where we use the temporal operators and assertions of the kinds:

- ∎ "*A* **U** *B*" to denote "*A* holds until *B* where *B* must eventually hold",

- "•*A*" to denote "*A* holds in the previous state", and
- "D*A*" to denote "*A* always holds".

Detailed formal definitions of linear-time temporal logic can be found in [BK08, Str06].

**Step 4** *Refine, Classify and Assess Hazards* — Guiding Questions: *Which failures are hazards? Which hazards are failures? Which hazards are no failures? Which hazards are relevant to be treated?* This step involves the investigation of hazards and their relationship to defects and nominal operation of the system. Starting from a more abstract model, lower-level *actions* and *modes* of the leaf *usage functions* or *environment tactics* are analyzed to extract defective parts of *operations*. This leads to refined assertions of hazards, more precise *classification* as well as a better quantification and determination of their *relevance*.

**Step 5** *Specify Safety Goals* — Guiding Questions: *How and where to allocate behavioral responsibility?* This step involves the transformation of identified hazards into goal assertions. One way how to do this has already been demonstrated in Section 3. In [Gle13], the smoothening step of using controlled natural language is left out. Instead, a direct transition to the use of temporal logic takes place. The safety goals can then be used as guarantee specifications for the respective usage functions or as assumption specifications for the respective environment tactics.

**Step 6** *Design the Safety Functionality* — Guiding Questions: *How does the functional safety concept have to look like?* Strategies for hazard mitigation deal with the question of how the system or the environment can be equipped to detect hazards and take over the control to avoid or mitigate impacts. This results in the introduction of *fail-safe transitions* and *passive* or *preventive transitions*. The implementation of such transitions usually amounts to *state observation or runtime diagnosis*.

After the last step the specification is transformed in a way to reduce hazards, to assess and to mitigate hazardous weaknesses.

## 4.3 Case Study: Application of the Procedure

The procedure of Figure 4.1 is now exemplified (further details in [Gle13]). The system is a commercial road vehicle ("truck" for short) and the environment the part of the world including the driver, a truck is usually performing in.

**Step 1** The *driving missions* of a truck and possible tactics of its driver are provided as use cases (Tables 4.1, 4.2 and 4.3). As physical interface phenomena, consider two vectors for speed, $v_I$ and $v_{oo}$, and two for position, $pos_I$ and $pos_{oo}$. Figure 4.2 shows the function hierarchy of a truck. The tactics are constructed similarly and shown in Figure 4.3. The upper levels of the hierarchy help identifying complex operations observable at the interface of a truck by co-executing its functions and the environment's tactics. Let us consider the operation move. At leaf level, Figure 4.4a shows the usage function $f^I_{use.StopBrake}$.

| UC #27 | Use truck (usage goal #27, $f^E_{Missions} \| f^I_{Truck}$) |
|---|---|
| *Scope* | $\mathcal{I}$; level: primary task in $f^*_{use}$; primary actor: $E$ |
| *Preconditions* | Enough fuel, battery on, etc. |
| *Minimal Guarantees* | Neither the trucker, his goods, nor the environment will be harmed. |
| *Success Guarantees* | The trucker accomplishes his mission by using the truck. |
| *Trigger* | The trucker activates the vehicle by applying the key. |
| *Description* (list of interaction descriptions) | 1. The trucker activates the vehicle by applying the key.<br>2. She performs, e.g., UC #5,10 to accomplish her missions.<br>3. The vehicle reacts properly to her commands.<br>4. The trucker deactivates the vehicle. |

Table 4.1: Use case #27 "Use truck"

| UC #5 | Park at steep hill (usage goal #5, $f^E_{use.Park} \| f^I_{use.Drive/Move}$) |
|---|---|
| *Scope* | $\mathcal{I}$; level: primary task in $f^*_{use}$; primary actor: $E$ |
| *Preconditions* | The truck is driving near a free and proper parking lot. |
| *Minimal Guarantees* | |
| *Success Guarantees* | The truck is parked in a parking lot at a steep hill compatible to the current mission goal. |
| *Trigger* | The trucker stops in front of a parking lot at a steep hill. |
| *Description* (list of interaction descriptions) | 1. The trucker stops in front of a parking lot at a steep hill.<br>2. She uses gas pedal, steering wheel, clutch, gears, brakes (UC#10) and rear mirrors to place the truck into the lot. |

Table 4.2: Use case #5 "Park at steep hill"

As there was no documentation available in [ITK12], the three use cases have been reconstructed from domain knowledge being collected during the project mentioned in Section 1.3 as well as during other research projects in the automotive domain, in which the chair was involved throughout the last years. The same holds for the function hierarchy.

**Step 2** Figure 4.4b shows failure possibilities as a fragment $f^\mathcal{I}_{fail.StopBrake}$.

**Step 3** Search for impacts stemming from truck operations from Step 1. As modeled in Figure 4.5, the operation move is physically relevant because it affects $p_\mathcal{I}$ and $pbs_\mathcal{I}$. An impact of move is a *collision* defined as a combination of *too small distances* and *too high relative velocities* and, thus, represented by an approximating condition

$$\mu \equiv |p_\mathcal{I} - p_{oo}| \geq v_{ok} \land |pbs_\mathcal{I} - pbs_{oo}| \leq max_{x \in \{oo,\mathcal{I}\}}\{diameter_x\} \qquad (4.1)$$

The candidate hazard as a condition $\tilde{\chi}$ for hazardous performance of move would be tied to situations, such as, e.g., move triggered or altered without foreseen user operation (Table 4.4). In Figure 4.5, the safe refinement of the operation $\tau$ denotes all transitions fulfilling the predicate *userOperation*. Consider $\tilde{\chi} \equiv$ *"Unintended start of movement"* formalized as hazard assertion

$$\tilde{\chi} \equiv \neg(move \vee userOperation) \mathbf{U}\ move \tag{4.2}$$

| UC #10 | Use brakes (usage goal #10, $f^E_{Missions} \| f^I_{use.Accelerate/Brake}$) |
|---|---|
| *Scope* | $I$; level: primary task in $f^*_{use}$; primary actor: $E$ |
| *Preconditions* | None. |
| *Minimal Guarantees* | The truck is slowing down. |
| *Success Guarantees* | The truck is properly slowing down or coming to a stable halt. |
| *Trigger* | The trucker actuates the brake pedal. |
| *Description* (list of interaction descriptions) | 1. The trucker actuates the brake pedal.<br>2. The truck decreases its speed accordingly.<br>3. Optional: When the truck comes to a halt, the trucker decides to activate the stop brake. |

Table 4.3: Use case #10 "Use brakes" always included by UC#5

| Step 1 | | | Step 3 | Step 4 | | | | Step 5 |
|---|---|---|---|---|---|---|---|---|
| $o$ | $\mu$ | $\tilde{\chi}$ | **Short Description of Hazard** | S | A | G | W | IC |
| move | - | - | Unintended start of movement | 2 | 2 | 1 | 1 | high |
| move | - | - | Unintended change of direction | 2 | 2 | 1 | 1 | high |

Table 4.4: Hazard assessment

From UC #5 we know a relevant local situation $\sigma \equiv$ *"The truck is standing in a steep parking lot and the stop brake is activated."*

**Step 4** Assertion 4.2 is refined by $move \equiv \mathit{v}_I /= 0$ and $userOperation \equiv gasPedal = pressed \wedge \exists x.(\bullet gearLever = x) \rightarrow gearLever /= x$ (Figure 4.5). The decomposition of the usage function $f^I_{use.Drive/Move}$ containing move into actions at the leaf level of the function hierarchy shows that, e.g., the mode **active** and the action **brake** of $f^I_{use.StopBrake}$ are physically relevant, because in this mode, this action is responsible to maintain $\mathit{v}_I = 0$ or to contribute to the operation **park**. The failure transition **suppressBrake** of $f^I_{fail.StopBrake}$ (Figure 4.4b) contributes to $\tilde{\chi}$ and potentially to impact $\mu$, e.g., because of gravity in $\sigma$, no driver input needed to reach move and the driver possibly not in the truck to take over control. This analysis leads to the assertion

$$\chi \equiv (\neg move \wedge mode(f^I_{StopBrake}) = active)\ \mathbf{U}\ move \tag{4.3}$$

**Step 5** Our safety goal for the truck implies $\neg \mu \equiv$ "no collision" which is broken down by:

$$\tilde{\gamma} \equiv \mathrm{D}(\neg move \rightarrow (\neg move\ \mathbf{U}\ (userOperation \wedge \neg move))) \tag{4.4}$$

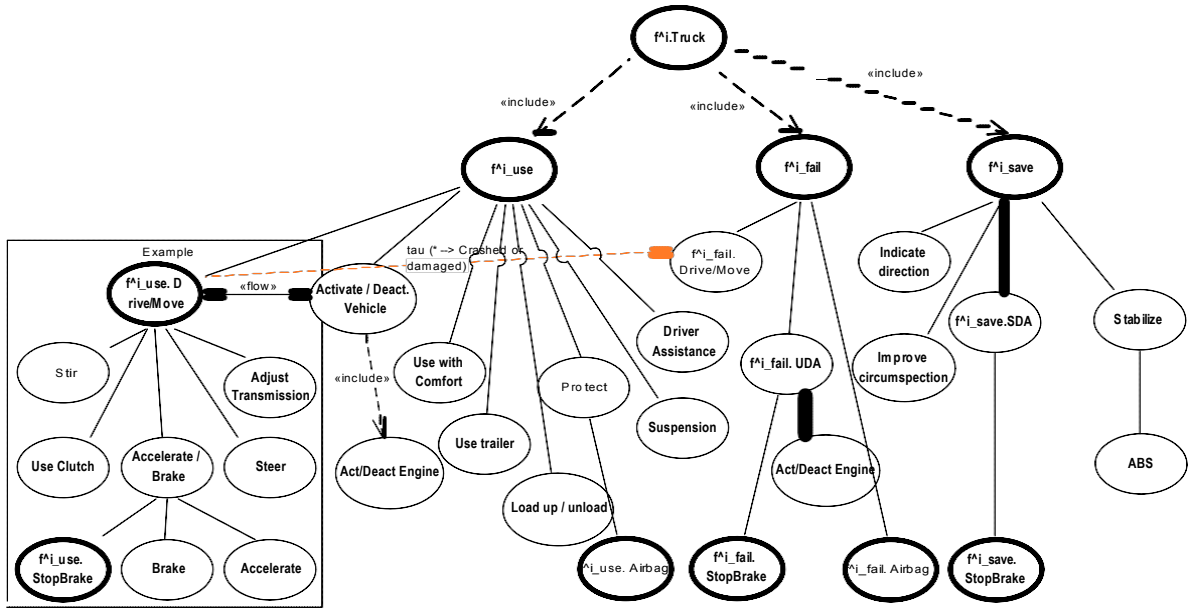

Figure 4.2: Excerpt of a truck usage function hierarchy

$\tilde{\gamma}$ is assigned to $f^{I}_{Drive/Move}$. From the list of relevant hazards like $\chi$ (step 4), safety requirements for $f^{I}_{StopBrake}$ have been derived by assigning a high integrity class.

**Step 6** Analyses, such as described in Section 5 or [SF11], provide more detailed characteristics of the physical action **suppressBrake**. To realize $f^{I}_{StopBrake}$, it has to be designed to mitigate or avoid the hazards $\chi$. The fail-safe transition in $f^{I}_{save.StopBrake}$ (Figure 4.6a) could incorporate a fail-silent mechanism suited to quickly mask potential **suppressBrake** transitions.

## 4.4 Summary

First, we used a systematic way to perform behavioral system specification using use cases and hierarchically arranged state machines. Based on the resulting system model, we tied defect knowledge to specific modes and behaviors of the system. Step 2 avoids the implicit assumption of the equivalence of failures and hazards as it was made in the first perspective (Section 3). This separates hazard analysis from reliability analysis to a certain extent. Using a full-fledged state machine model is an enhancement of the results shown in Section 3, as the state machine represents *many or even all relevant system runs* in a quite compact manner. It can be used as a *behavioral approximation of the considered system* to be checked against the hazards and safety properties derived in the first perspective but also in Steps 3 and 4 of this perspective. We sometimes use predicates and terms in our temporal logic assertions. For automation, the formulas still need to be abstracted. However, this is not discussed in detail. Furthermore, based on the same state machine and the results of property checking, we sketched how measures can be specified as refined or superimposed transitions for the treatment, i.e., avoidance or mitigation, of hazards.

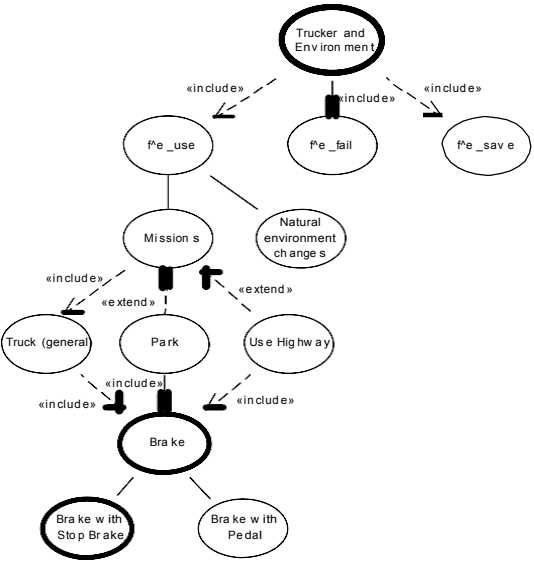

Figure 4.3: Excerpt of an environment tactics hierarchy

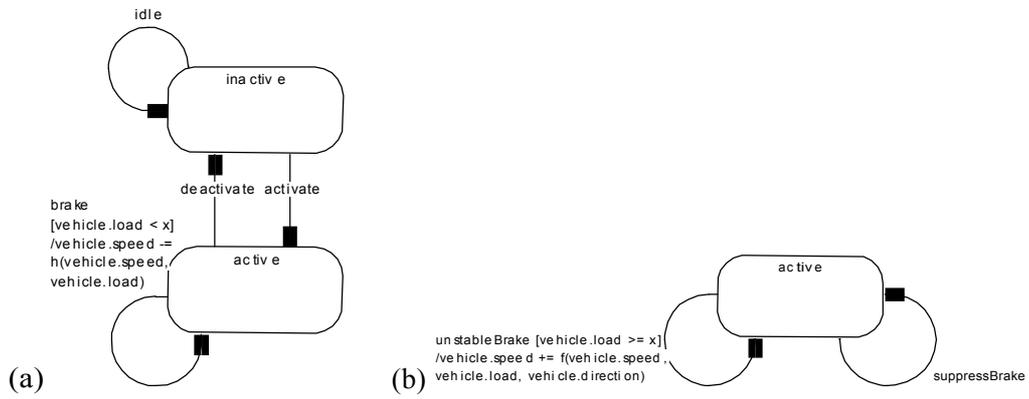

Figure 4.4: The usage function $f^{\mathcal{I}}_{\text{use.StopBrake}}$ (a) and possible defects $f^{\mathcal{I}}_{\text{fail.StopBrake}}$ (b)

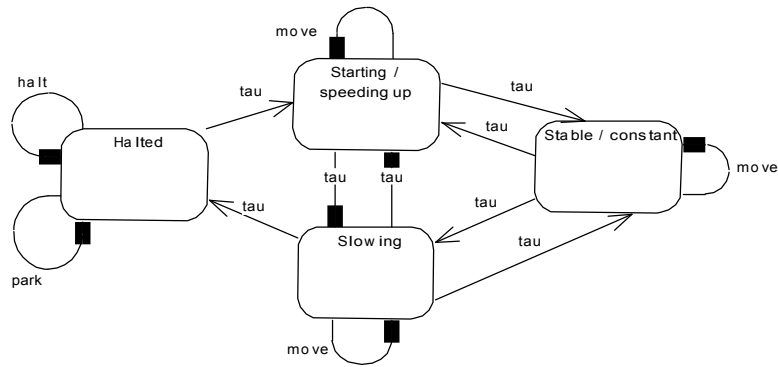

Figure 4.5: The usage function $f^{\mathcal{I}}_{\text{use.Drive/Move}}$

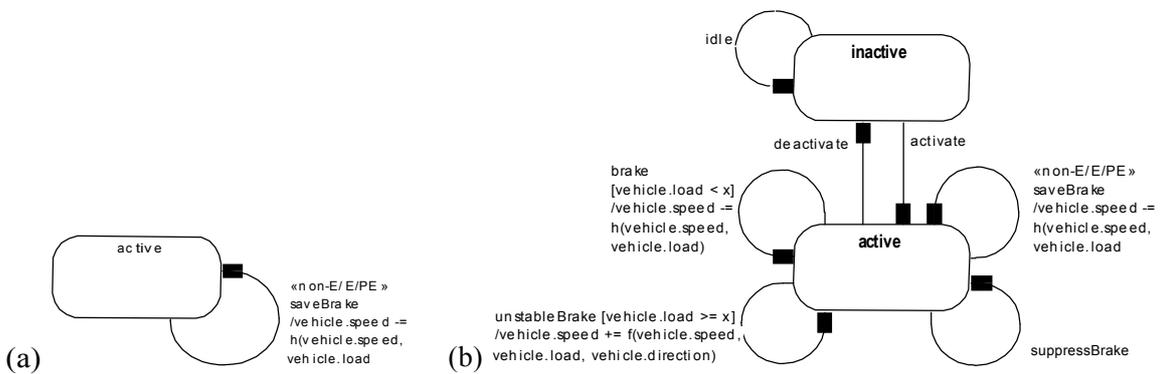

Figure 4.6: The safety fragment $f^{\mathcal{I}}_{\text{save.StopBrake}}$ (a) and the overall function $f^{\mathcal{I}}_{\text{StopBrake}}$ (b)

# 5 Automated Safety Analysis based on Component-oriented Modeling

Safety analysis may have to be carried out repetitively for different versions and variants during the design of a system, is knowledge-intensive, and consumes significant efforts of experts. The objective of providing powerful tools for supporting, or even automating, a major part of the safety analysis process is a challenge to artificial intelligence (AI) approaches, since it involves reasoning about the physical world.

Our third perspective presented in this chapter presents a general AI approach to safety analysis, a tool for automated generation of safety analysis results as well as its application and evaluation. Since a modern vehicle is an aggregation of different subsystems that are controlled by software and that interact with each other and a dynamic environment, addressing this task can be seen as a contribution to the more general, important problem of safety of cyber-physical systems.

## 5.1 Key Ideas

Our solution to the case study on the drive train of a truck (Section 2) exploits qualitative modeling and a qualitative spatial representation. This reflects the division of the problem and model into two parts, namely modeling and inferring

- **abnormal behavior of the vehicle** (called hazards) as a consequence of component faults and

- the impact of a hazard **on its environment**.

The nature of the **worst-case analysis** (determining qualitative effects of classes of component faults under abstract classes of scenarios) enables, and even stronger **requires**, the use of **qualitative** representations and models.

For instance, in our case study, qualitative behavior models of the components of the drive train are used to predict the effect of a component fault on the motion of the entire vehicle, e. g., an unintended acceleration. The analysis of the impact of this effect determines potential collisions due to the disturbed motion of the vehicle based on a qualitative spatial representation of positions of the vehicle and other objects relative to the road and their interference for different abstract scenarios.

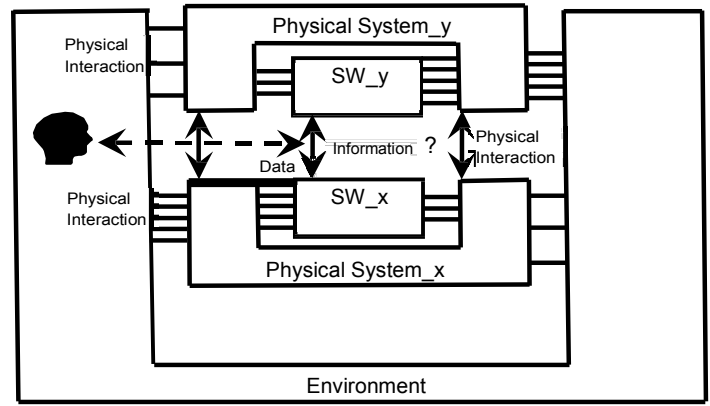

Figure 5.1: The vehicle as an aggregation of physical and software components in an environment

## 5.2 Requirements and Approach

The current practice and standards do not directly lend themselves to precise definitions of the involved concepts and types of required inferences, nor do they imply particular ways of structuring models and processes (Section 1.2). We do so following the perspective illustrated by Figure 5.1: a cyber-physical system comprises a number of subsystems, which are systems composed of physical (e.g., mechanical, electrical, hydraulic) components and software components, whose interaction happens exclusively through a usually relatively small set of sensor signals as an input to and actuator signals as an output of the software component(s). Different subsystems interact both via connections between their physical components and via communication between their software components. In a vehicle, the components of the drive train with their individual ECUs are examples for such subsystems. At a higher level, the drive train itself can be considered as a subsystem as shown in the first two perspectives (Sections 3 and 4). The top-level system is the entire vehicle.

From the point of view of safety analysis, it is important to note that it is **solely** the **physical** system, i. e., the vehicle (or its physical parts) that **interacts with the environment**. The embedded software never directly interferes with the environment. As a consequence, hazards, misbehaviors that bear the potential of damage in the environment, are defined exclusively at the intersection of the physical system and the (physical) environment. Even unexpected operations carried out by the software – are never a hazard per se. They may only cause one via the response of the physical system to the actuator signals.

As an important consequence, faulty software behavior matters if and only if it may cause the physical system to create a hazard. Therefore, our approach moves the (model of the) **physical system into the center** and models software – and especially software faults – solely with regard to the physical model.

On the other hand, hazards create risks only through their **impact on the environment**, which includes other agents or objects. Obviously, this environment is much more diver-sified and dynamically changing compared to the designed artifact, the vehicle. It cannot be explored exhaustively, but only through certain abstract types of scenarios and driving situation.

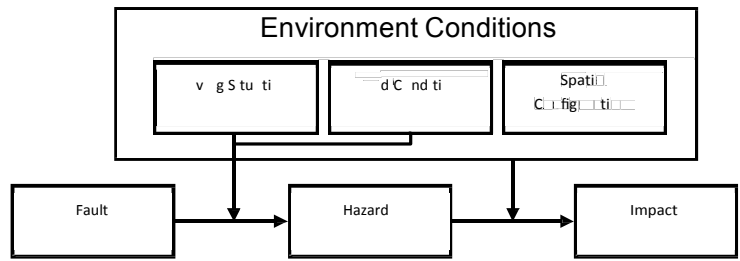

Figure 5.2: Core elements of automated hazard and impact analysis

In consequence, we approach the task of building a tool for safety analysis by dividing it (conceptually) into two steps (Figure 5.2):

- **hazard analysis**: a model of the vehicle is used to determine whether assumed faults of (software or physical) components may result in (pre-defined) hazards for a set of specified driving conditions (in terms of speed, driver actions, etc.) and road condition (road surface, slope).

- **impact analysis**: a model of the environment, which includes the vehicle as well as other objects and agents, determines whether the hazard may have a dangerous impact (in our case, a collision) under certain environment conditions, which include the driving and road conditions (e. g., curves) and the spatial configuration of other objects.

The two models together, associate component faults with their impacts, i.e., safety violations and risks. Based on the composed model, the automated analysis (Figure 5.2) can be carried out in one step, directly relating component faults and impacts, without the intermediate effects, the hazards.

We mention that, based on the above statements, we currently exclude two relevant aspects: Firstly, software faults may indirectly create a safety critical situation, e. g., by supplying the driver with wrong, too little, or too much information and, thus, causing an inappropriate driver action. Secondly, we do not consider the physical impact of hazards on the driver and passengers of the vehicle.

Finally, we derive some design decisions from requirements and the inherent nature of the task. As emphasized before, the analysis is highly repetitive, demanding, and has to be performed several times during the design phase, applied to alternative designs, and subsequently to different versions and variants. This does not necessarily make the manual analysis more reliable but more error prone. Any proposed solution to supporting the process will only be economically beneficial if it does not multiply efforts along with the repetition, as well. More specifically, if for each analysis, a model of the respective subsystem variant has to be built from scratch, this is unlikely to be superior to a manual analysis. The answer to this challenge lies in the **reuse of models**, i. e., **compositional modeling**, where system models are composed from component models in a library.

The nature of the analysis makes compositional modeling feasible: it is an inherently **qualitative** and a worst-case analysis. Firstly, the analysis is performed at design time, and parameters may not yet have numerical values. Beyond this, the faults are qualitative: decreased friction of a brake, a leakage of a pipe, a high sensor signal etc. cannot be described by numerical values, but only by ranges of them. Hazards are qualitative: too high or too low acceleration, and best specified at this level of granularity. Environment conditions are qualitative: "a vehicle approaching a pedestrian crossing with medium speed" or "going downhill a winding road". With regard to the required inferences, the worst-case analysis is not expected (and, given the qualitative input, not able) to firmly conclude the impact. What needs to be determined is the **potential** of a collision, e.g. given a reduced deceleration of the vehicle and, hence, a longer brake path – considering that it is uncertain whether pedestrians are present or not. Nevertheless, determining that a brake with reduced friction results in a reduced deceleration suffices to consider it as a reason for a risk in the respective scenario.

Hence, we need qualitative models and representations and inferences determining the possibility of hazards and severe impacts. At this point, we note that the qualitative nature of the required models provides the basis for reusable models and cheap model building. The impact on the level of abstraction of the models will be shown in the following sections.

## 5.3 A Formalism for Qualitative Deviation Models

As motivated by the discussion above (Section 5.2), the characteristics of the models used in our solution are the following:

- **Compositional** modeling: models of systems are obtained through aggregation of models of its parts, possibly across several layers of hierarchy.

- **Component**-oriented modeling: the parts are components, i.e. the building blocks that are assembled to form the system and determine its behavior (both physical and software components). This is due to two reasons. Firstly, component models can be reused in different system models just as the components are reused in different systems. Secondly, components are the entities that are subject to faults, whose impact needs to be determined in safety analysis.

- **Qualitative** behavior models reflect the nature of the analysis, as pointed out above.

- **Relational** models (as opposed to functions or transition systems) are chosen to represent these qualitative behavior descriptions, based on the observation that hazards are commonly the result of a fault in one state of the physical system (rather than occurring after a sequence of state transitions).

- **Deviation** models are used, since faults, hazards, and impact are characterized as (qualitatively) distinct from nominal behavior.

In the following, we specify these characteristics more formally, though in a nutshell (for introductory material, see [Str97, Str00, SP03, HdKC92, Str08]).

**Component-oriented Modeling** A component type (used to create different instances) is represented under a structural and a behavioral perspective:

It has a number of typed **terminals**, which can be shared with other components.

Thus, a **system structure** is described as $(COMPS, CONNECTIONS)$, where $COMPS$ is a set of (typed) components and $CONNECTIONS$ is a set of pairs of terminals of equal type belonging to different components.

A component $C_i$ has a vector $\vec{v}_i = (v_{ik})$ of **variables**, comprising **parameters** and **state variables**, which are considered as internal and constant and dynamically changing, resp., and **terminal variables**. The latter are obtained as instances of terminal types which have a set of associated variable types.

The $CONNECTIONS$ of a system structure induce a set $VARIABLECONNECTIONS$ of pairs of corresponding terminal variables from connected components. Each variable connection introduces a mapping between the values of the connected variables: this is usually equality (for signals, voltage etc.), while for directed variables, such as torque and current, the sign is flipped.

A component $C_i$ has a set of **behavior modes** $\{mode_j(C_i)\}$, where one mode, $OK$, corresponds to the nominal behavior of the component and the other ones denote different defects of the component.

**Qualitative Modeling** Qualitative models describe component behavior in terms of variable **domains** $DOM(v_{ik})$ that are **finite**. Besides domains that are considered "naturally" discrete, such as Boolean for binary signals and $\{OPEN, CLOSED\}$ for the state of a clutch, the domains of continuous variables are obtained by discretization and are usually **finite set of intervals** that reflect the essential distinctions needed for capturing the relevant aspects of component behavior:

$$DOM(v_{ik}) = \{I_{ikm} | m = 1, 2, \ldots, n\}.$$

**Relational Modeling** The behavior of a component under a particular behavior mode, $mode_j(C_i)$, is represented as the set of qualitative tuples that are possible if this mode is present, i.e., as a **relation**

$$R_{ij} \subset DOM(\vec{v}_i) = DOM(v_{i1}) \times DOM(v_{i2}) \times \cdots \times DOM(v_{ir}),$$

or, in AI terminology, as a **constraint** (which means many operations on models introduced in the following can be realized using techniques of Finite Constraint Satisfaction). Each variable connection $(v_p, v_q)$ introduces a connection relation $R_{pq}$ capturing the mapping between domains.

**Compositional Modeling**  A model of an aggregate system is not unique, but dependent on the behavior modes of the components. A mode assignment $MA = \{mode_j(C_i)\}$ specifies a unique behavior mode for each component, and a model of the system is obtained as the (natural) join (as in the relational algebra and SQL, see [Cod70]) of the mode model relations and the connection relations:

$$R_{MA} = (\bowtie R_{pq}) \bowtie (\bowtie R_{ij}) \tag{5.1}$$

**Deviation Models**  Some faults are stated in absolute terms ("zero braking torque exerted by brake"), while others are only described in relative terms ("reduced braking torque produced by a worn brake"), and so are definitions of hazards: "reduced deceleration of vehicle". Such models are meant to capture qualitative deviations from the nominal behavior, which is the basis for detecting deviations in the behavior of the entire system. We use deviation models in the same way as [Str04], [SF12a]: the qualitative deviation of a variable $x$ is defined as

$$\Delta x := sign(x_{act} - x_{nom}), \tag{5.2}$$

which captures whether an actual (observed, assumed, or inferred) value is greater, less or equal to the nominal value. The latter is the value to be expected under nominal behavior, technically: the value resulting when all components are in *OK* mode. Qualitative deviation models can be obtained from standard models stated in terms of (differential) equations by canonical transformations, such as

$$a + b = c \Rightarrow \Delta a \oplus \Delta b = \Delta c$$

$$a * b = c \Rightarrow (a_{act} \otimes \Delta b) \oplus (b_{act} \otimes \Delta a) \ominus (\Delta a \otimes \Delta b) = \Delta c.$$

Here, $\oplus, \otimes, \ominus$ are addition, multiplication, and subtraction operators of **interval arithmetic** applied to signs represented as $(-\infty, 0), [0, 0], (0, \infty)$.

## 5.4 Automated Prediction of Effects

From a logical point of view, the two steps in the process of Figure 5.2 are similar:

- **Hazard analysis** requires checking whether a **hazard** is, or may be, **caused** by an assumed **fault** under given driving conditions. Logically, this means determining whether the hazard is **implied by or, at least consistent with**, the model of the faulty vehicle (with the respective fault model included) and information about the driving situation and road condition.

- **Impact analysis** requires the same check for the **impact**: is it **implied by or, at least consistent with**, the **model of the faulty vehicle and information about the environment conditions**.

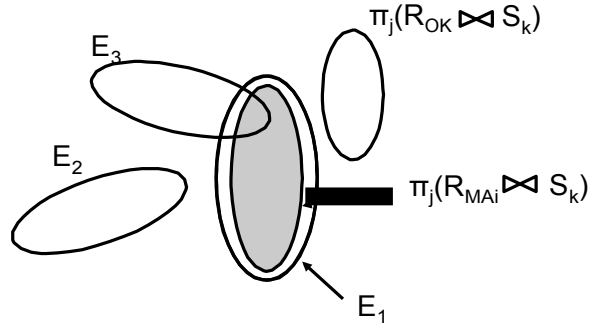

Figure 5.3: A relational perspective on determining effects based on fault modes and scenarios $S_k$

Actually, hazard analysis is equivalent to FMEA restricted to system-level effects, since FMEA analyzes whether or not a component fault leads to a predefined effect (which is a violation of the intended functionality and safety) under a certain scenario (mission or mission phase). This is why, for both hazard analysis and the overall impact analysis, we exploit an algorithm that has been used for FMEA [PCB$^+$04, SF12a]. The algorithm is based on representing not only behavior models as finite relations (as described in Section 5.3), but also effects and scenarios. Effects can naturally be stated as relations $E_j$ on system variables that characterize the relevant aspects of system behavior, such as (the deviation of) the acceleration of a vehicle, while a scenario is typically a relation $S_k$ on exogenous variables and internal states of the system like the position of the brake pedal (pushed or not) and the vehicle speed. The algorithm checks the presence of effects for each possible single fault in the system under each defined scenario. Using the relational representation, this means that for a mode assignment $MA_i$ that contains exactly one fault mode and $OK$ modes otherwise, the respective behavior model $R_{MA_i}$ is automatically composed according to Equation 5.1. Then, for each scenario $S_k$ and each effect $E_j$, it is determined whether

- $\pi_j(R_{MA_i} \bowtie S_k) \subseteq E_j$,
  where $\pi_j$ denotes the *projection* (as used in relational algebra [Cod70]) to the variables of $E_j$. The positive case, i.e., the faulty behavior is included in the effect, means that the effect will **definitely occur** (case $E_1$ in Figure 5.3). Stated in logic, this means that the fault entails the effect in this scenario.

- $\pi_j(R_{MA_i} \bowtie S_k) \cap E_j = \emptyset$.
  If the intersection is empty, the effect does not occur (case $E_2$). Logically, the effect is inconsistent with the fault mode and the scenario.

- Otherwise, the effect **possibly occurs** (case $E_3$), i.e., $R_{MA_i} \bowtie S_k$ covers both conditions under which the effect is present and others under which it does not occur – the effect is consistent with the fault mode and the scenario.

Figure 5.3 also indicates that all effects should be disjoint from $\pi_j(R_{OK} \bowtie S_k)$, where $R_{OK}$ is the model for the mode assignment of $OK$ to all components. If this is not fulfilled, this would indicate a design fault or an improper effect definition (because the correct device may violate the requirements) – or, of course, a modeling bug.

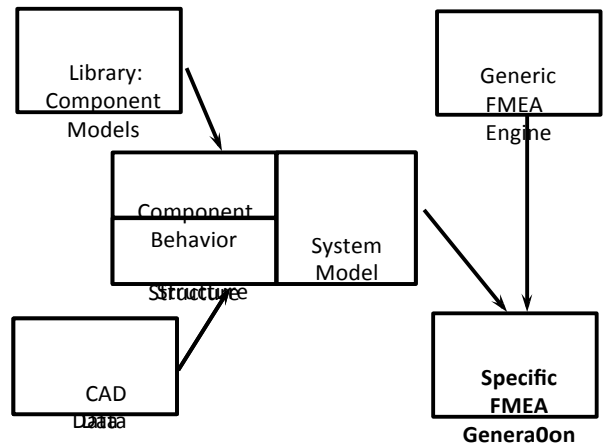

Figure 5.4: Automatic generation of a specific FMEA analysis by automated model composition and application of a generic inference engine.

The model based systems tool Raz'r [OCC95] includes an implementation of an algorithm that performs the checks specified above for each scenario and fault and generates the FMEA table. Exploiting compositional modeling, also the generation of the model of the (faulty) system is automatic, such that an FMEA tailored to a specific device requires solely a description of the device structure as an input, as indicated in Figure 5.4.

We will employ the algorithm in three ways, namely for deriving

- **Hazards from fault modes**, where the fault modes are components faults, and scenarios are the driving situations and road conditions (Section 5.5)

- **Impacts from hazards**, using the hazards as the fault modes and specifying scenarios as environment conditions (Section 5.6)

- **Impacts from fault modes**, which are component faults and analyzed under different environment conditions (Section 5.6.4).

## 5.5 Case Study: Model-based Hazard Analysis

Based on Section 2, this section gives details on the modeling of each component of the case study and applies part of the reasoning procedure explained above. Throughout this section, most variables have values from the domain $Sign = \{-, 0, +\}$: torques (T) and forces (F), rotational and translational speeds ($\omega$ and $v$) and translational acceleration $a$, $\omega$ and $v$, together with their deviations, in which case the $\Delta$ symbol precedes the above mentioned variables. The commands and states explicitly discussed here have Boolean values $\{0, 1\}$.

### 5.5.1 Modeling the Drive Train

The core purpose of the drive train component models is to determine the (deviation of the) torque acting on the wheels, which determines the (deviation of the) translational acceleration of the vehicle (if the road surface permits).

Faults may introduce non-zero deviations, e.g. the model of a worn brake would generate a deviating braking torque, which depends on the direction of a non-zero rotation (static friction)

$$\Delta T_{brake} = \omega \tag{5.3}$$

or in the direction of the applied torque in case of kinetic friction

$$\Delta T_{brake} = T_{wheel} \tag{5.4}$$

Models of both, *OK* behavior and faulty behavior, are stated in terms of constraints (i.e., relations) on the deviations. For instance, a closed clutch propagates a deviating torque arriving on the left (i.e., of the engine) further to the right (under a change of the sign):

$$\Delta T_{right} = -\Delta T_{left}. \tag{5.5}$$

The overall torque is not determined locally, but results from the interaction of all mechanical components. The engine can produce a driving torque, the braking elements (wheel brake, retarder, engine) may generate a torque opposite to the rotation, and the clutch and transmission may interrupt or reverse the propagated torque.

Our current model is based on assuming that there are no cyclic structures among the mechanically connected components, which is the case in our application, but certainly also in a much broader class of systems. The component models link the torque (deviations) on the right-hand side to the one on the left-hand side, possibly adding a torque (deviation) generated by the respective component. Hence, at each location in the drive-train model, the torque represents the sum of all torques collected left of it, and so does the torque deviation.

Whenever a component terminates the torque propagation (e.g., the wheel or an open clutch), the arriving torque is also the total one for the section left. For an open component, the torque on the right-hand side is zero, as exemplified by the clutch ($state = 0$ means open):

$$state = 1 \Rightarrow T_{right} = T_{left} \tag{5.6}$$
$$state = 0 \Rightarrow T_{total} = T_{left} \land T_{right} = 0 \tag{5.7}$$

Determining the deviation models is not as straightforward as it may appear, as we will explain using the model of the retarder as an example. If engaged ($state = 1$), it will generate a torque $T_{brake}$ opposite to the rotation (zero, if there is no rotation) and add it to the left-hand one. The base model is obvious:

$$T_{right} = T_{left} \oplus T_{brake} \tag{5.8}$$

| cmd | $\Delta cmd$ | $\Delta state$ |
|-----|--------------|----------------|
| 1   | 0            | 0              |
| 0   | 0            | +              |
| 0   | -            | 0              |
| 1   | +            | +              |

Table 5.1: Retarder stuck engaged - Deviation constraint

$$state = 1 \Rightarrow T_{brake} = -\omega \quad (5.9)$$

$$state = 0 \Rightarrow T_{brake} = 0 \quad (5.10)$$

where $\oplus$ denotes addition of signs. The first line directly translates into a constraint on the deviations:

$$\Delta T_{right} = \Delta T_{left} \oplus \Delta T_{brake} \quad (5.11)$$

However, determining $\Delta T_{brake}$ requires consideration of how the actual state is related to the nominal one, which depends on the control command to the component, and, to complicate matters, not on the actual command, but the **command that corresponds to the nominal state**. This means we have to model possibly deviating commands, and we apply the concept and definition of a deviation also to Boolean variables. For instance, in the retarder model, $\Delta state = -$ means $state = 0$ (i.e., it is not engaged) although it should be 1, and $\Delta state = +$ expresses that it is erroneously engaged. Such deviations could be caused by retarder faults, e.g. stuck-engaged. However, in the context of our analysis, we must consider the possibility that the commands to the retarder are not the nominal ones (caused by a software fault or the response of the correct software to a deviating sensor value). Under multiple faults, a component fault may even mask the effect of a wrong command (the retarder stuck engaged compensates for $\Delta cmd = -$). In the OK model of the retarder, the component does what the command requests and the deviations of the command and state (i.e., the real, physical state) are identical:

$$\Delta state = \Delta cmd \quad (5.12)$$

Table 5.1 captures the constraint on the deviations for a fault "stuck engaged". Here, the third row represents the masking case mentioned above, while the first row reflects that the physical state coincides with the command, while in the second row, it does not. From $\Delta state$, $\Delta brake$ is determined by

$$\Delta T_{brake} = -\omega \otimes \Delta state \quad (5.13)$$

where $\otimes$ denotes multiplication of signs. This completes the model of the retarder.

### 5.5.2 Software Models

Since the drive train contains a number of ECUs, we also need to include models of software in our library. More specifically, we also have to include models of software faults. In principle, the space of software faults is infinite and enumerating and modeling faults in our context may seem infeasible. However, the fault models we need do not have to capture a detailed description of what can be bugs in the ECU software (missing termination criteria, wrong thresholds, omitted statements, etc.), which would lead to a huge set of fault models. Remember: all that matters about software faults is their impact on the physical system, more precisely, on the controlled actuators. For the Boolean commands in our model, this means the only fault types to be considered are

- Missing (or late) command: $\Delta cmd = -$
- Untimely (or early) command: $\Delta cmd = +$.

The same applies to continuous actuator signals, where the faults represent "signal too low" and "signal too high", respectively. This illustrates our claim that putting safety analysis back on its feet and the physical model in the center, greatly simplifies the modeling and analysis of the embedded software. In particular, for the purpose of hazard analysis, we obtain a small set of reusable software models for our library. Of course, if we do have a more detailed model of the software, also the fault models can be more specific.

### 5.5.3 The Model Library

Along these lines, a library of component type models has been produced that covers the basic mechanical components of the drive train and the related ECUs:

- Engine
- Crankshaft
- Clutch
- Gear box
- Retarder
- Wheel
- Wheel brake
- Engine ECU
- Transmission ECU (controlling the gear box and the clutch)
- Retarder ECU
- Brake ECU

Furthermore, there are some special components, whose interaction with the other components contributes to the behavior of the entire system:

- **Load**, a virtual "component" that represents the mass of the entire truck. It contributes to determining the gravitational force and transforms the force acting on it to acceleration
- **Road**, whose friction influences how translational forces and velocities relate to torques and rotational speeds and whose slope determines the gravitational force

5 Automated Safety Analysis based on Component-oriented Modeling

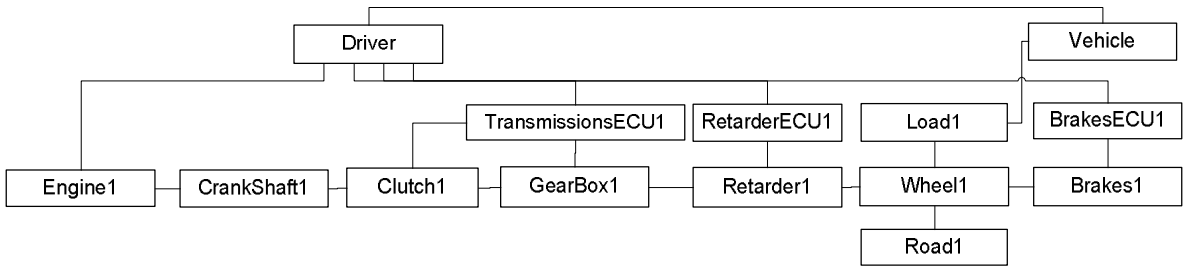

Figure 5.5: Vehicle model (Screenshot from Raz'r [OCC95])

- **Driver**, creating driving conditions by pushing the accelerator or the brake pedal, selecting gears, etc.

- **Vehicle**, the entire system, capturing the behavior (deviation) of the truck, currently solely the speed and acceleration deviation, but in future extensions also the steering angle (deviation). Therefore, it represents the link to the environment model, as will be shown later.

From this library, the vehicle model is generated automatically in Raz'r [OCC95] from a structural system description, which is constructed by drag and drop in a CAD-like manner, as depicted in Figure 5.5.

### 5.5.4 Driving Situations and Road Conditions

The vehicle model (Figure 5.5) is used to predict the potential misbehavior of the vehicle in the presence of different component faults and under different conditions. As indicated in Figure 5.2, we split these conditions into

- driving situation, which characterizes the state of the vehicle

- road condition, i. e., state variables friction and slope of the "component" Road introduced above.

The list of driving situations (see Table 5.2) we consider is a set of common plausible and technically feasible combinations of **driver inputs** to the vehicle and the **speed**. For instance, the accelerator implies the respective signal to the engine ECU. The basic situations are: starting, driving, and braking in both forward and backward direction. For driving and braking in forward direction, we distinguish between two orders of magnitude of speed, low (+) and high (++). This distinction is actually not relevant to determining deviations in acceleration, but influences the impact analysis, because it results, for instance, in different brake distances.

Implausible combinations like pushing accelerator and brake pedal at the same time are omitted here (although the model covers this condition). Some plausible ones are currently not included, such as pushing the brake pedal at speed zero. This is of interest when stopping on a slope which we have not considered in this report.

| Driving situation | Accelerator pushed | Brake pedal pushed | Chosen gear | | | Clutch pedal not pushed | $v$ |
|---|---|---|---|---|---|---|---|
| | | | F | N | R | | |
| R-accelerate | x | | | | x | X | - |
| R-start | x | | | | x | X | 0 |
| F-start | x | | x | | | X | 0 |
| Drive high speed | x | | x | | | X | ++ |
| Drive low speed | x | | x | | | X | + |
| R-brake | | x | | | X | X | - |
| Stop | | x | X | | | X | 0 |
| F-brake high speed | | x | X | | | (X) | ++ |
| F-brake low speed | | x | X | | | (X) | + |

Table 5.2: Definition of driving situations

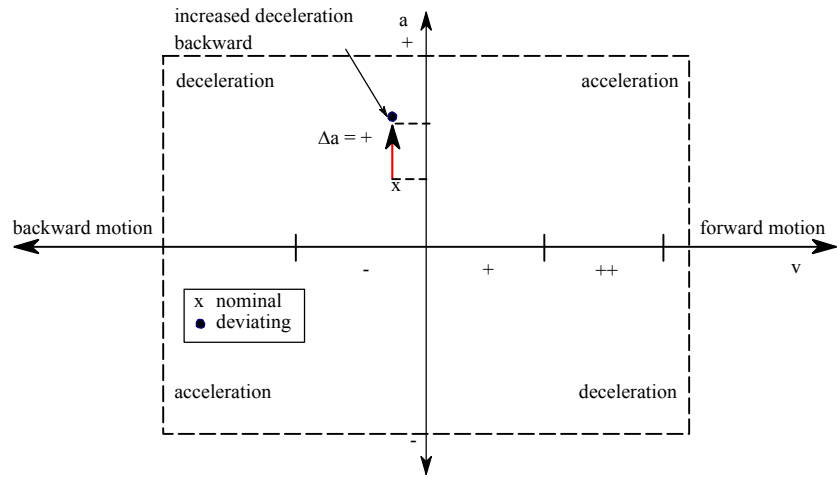

Figure 5.6: Qualitative modeling: location of nominal and deviating accelerations exemplified for one hazard and one driving situation (ID 13 in Table 5.3)

The road conditions in terms of surface friction and slope have, so far, been fixed to + (sufficient friction for preventing sliding or free spinning of the wheel) and 0 (horizontal road), although the models include the other cases.

### 5.5.5 Hazard Definition

The hazards (corresponding to effects in FMEA) are given by **deviating accelerations** acting on the vehicle, $\Delta a \in \{-, +\}$. Hence, basically, there are only two hazards, reduced or increased acceleration (relative to forward direction). However, an intuitive physical interpretation reflects various driving situations with different (intended) directions of the motion/acceleration. For instance, $\Delta a = +$ in forward driving means increased acceleration and higher speed than intended; but for braking in backward motion (i. e., velocity $v = -$, acceleration

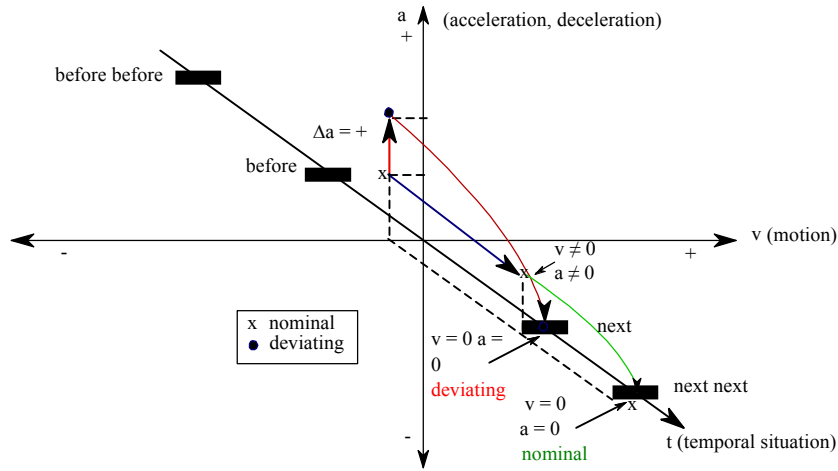

Figure 5.7: Location of nominal and deviating accelerations exemplified for one hazard and one driving situation (ID 13 in Table 5.3) in the qualitative time domain

$a = +$ and $\Delta a = +$) it states that the vehicle decelerates faster. See Figure 5.6, where this case is located in the top left quadrant. Hence, for presenting the results of hazard analysis, we introduce condition-dependent hazards according to Table 5.3 which extends and refines Table 2.1. The first and the last row in this table correspond to the two cases discussed above.

Figure 5.6 locates the nominal and deviating accelerations of the vehicle. Based on that, Figure 5.7 shows two possible **evolutions** using a qualitative time domain consisting of five temporal situations $t = \{before\text{-}before, before, now, next, next\text{-}next\}$: For increased deceleration (deviating acceleration), $v = 0$ already at time $t = next$, and for nominal deceleration (nominal acceleration), $v = 0$ not until time $t = next\text{-}next$. The diagram represents the case of a simplified linear physical motion over qualitative time. In a quantitative physical setting, we would have a vector equation of the form $\vec{p}(t+1) = \vec{p}(t) + \vec{a}(t) + \Delta \vec{a}(t)$ to capture the physical situation more precisely.

### 5.5.6 Results of Hazard Analysis

Hazard analysis was carried out using Raz'r's FMEA engine with the above specified driving situations (plus friction and slope fixed) as scenarios and the hazards as defined in Sections 5.5.4 and 5.5.5. In the Figures 5.8, 5.9 and 5.10, some of the automatically generated tables for the situations "F-start, F-brake, Drive" with forward motion are shown (where the distinction between high and low speed does not matter and is omitted). The uncolored rows are hazards that are definite, i.e., logically entailed by the respective fault and driving situation, while the **highlighted** rows indicate **possible** hazards (logically: consistent with fault and driving situation), according to the cases discussed in Section 5.4.

In the driving situation "Drive" (Figure 5.10), a retarder stuck in engaged position may overcome the accelerating torque ("unintended deceleration") or only diminish this quantity ("re-duced or no acceleration"). Both cases are actually possible, and the qualitative models correctly produce this ambiguous result. As a less obvious result, the clutch stuck open in the driving situation "F-brake" (Figure 5.9) triggers "Reduced or no deceleration", because it prevents the engine from contributing to the braking torque.

| ID | Hazard | Driving Situation | $a$ | $\Delta a$ |
|---|---|---|---|---|
| 1 | Increased acceleration | Drive, F-start | + | + |
| 2 | Reduced or no acceleration | Drive, F-start | +, 0 | - |
| 3 | Unintended deceleration | Drive | - | - |
| 4 | Unintended backward acceleration | F-start, R-brake | - | - |
| 5 | Reduced or no deceleration | F-brake | -, 0 | + |
| 6 | Increased deceleration | F-brake | - | - |
| 7 | Unintended acceleration | F-brake | + | + |
| 8 | Increased backward acceleration | R-accelerate, R-start | - | - |
| 9 | Reduced or no backward acceleration | R-accelerate, R-start | -, 0 | + |
| 10 | Unintended deceleration backward | R-accelerate | + | + |
| 11 | Unintended forward acceleration ID3) | R-start | + | + |
| 12 | Reduced or no deceleration backward | R-brake | +, 0 | - |
| 13 | Increased deceleration backward | R-brake | + | + |

Table 5.3: Definition of hazards, multiple values in a cell indicate a disjunction, extends and refines Table 2.1

| Scenario | Part | Failure Mode | Hazard / Impact |
|---|---|---|---|
| FstartSituation | CrankShaft1 | Broken | :Reduced_or_no_acceleration |
| FstartSituation | Clutch1 | ClutchStuckOpened | :Reduced_or_no_acceleration |
| FstartSituation | Clutch1 | ClutchStuckClosed | :>>no system level effects<< |
| FstartSituation | GearBox1 | StuckReverse | :Unintended_backward_acceleration |
| FstartSituation | GearBox1 | StuckNeutral | :Reduced_or_no_acceleration |
| FstartSituation | GearBox1 | StuckForward | :>>no system level effects<< |
| FstartSituation | Retarder1 | RetarderStuckNotEngaged | :>>no system level effects<< |
| FstartSituation | Retarder1 | RetarderStuckEngaged | :>>no system level effects<< |
| FstartSituation | Brakes1 | StuckNotEngaged | :>>no system level effects<< |
| FstartSituation | Brakes1 | StuckEngaged | :Reduced_or_no_acceleration |
| FstartSituation | BrakesECU1 | MissingCommand | :>>no system level effects<< |
| FstartSituation | BrakesECU1 | UntimelyCommand | :Reduced_or_no_acceleration |
| FstartSituation | RetarderECU1 | MissingCommand | :>>no system level effects<< |
| FstartSituation | RetarderECU1 | UntimelyCommand | :>>no system level effects<< |
| FstartSituation | TransmissionsECU1 | MisingClutchCommand | :Reduced_or_no_acceleration |
| FstartSituation | Engine1 | LowTorque | :Reduced_or_no_acceleration |
| FstartSituation | Engine1 | HighTorque | :Increased_acceleration |

Figure 5.8: Hazard analysis for the driving situation "F-start"

| Scenario | Part | Failure Mode | Hazard / Impact |
|---|---|---|---|
| FbrakeSituation | CrankShaft1 | Broken | :Reduced_or_no_deceleration |
| FbrakeSituation | Clutch1 | ClutchStuckOpened | :Reduced_or_no_deceleration |
| FbrakeSituation | Clutch1 | ClutchStuckClosed | :>>no system level effects<< |
| FbrakeSituation | GearBox1 | StuckReverse | |
| FbrakeSituation | GearBox1 | StuckReverse | :Reduced_or_no_deceleration |
| FbrakeSituation | GearBox1 | StuckReverse | :Unintended_acceleration |
| FbrakeSituation | GearBox1 | StuckNeutral | :Reduced_or_no_deceleration |
| FbrakeSituation | GearBox1 | StuckForward | :>>no system level effects<< |
| FbrakeSituation | Retarder1 | RetarderStuckNotEngaged | :Reduced_or_no_deceleration |
| FbrakeSituation | Retarder1 | RetarderStuckEngaged | :>>no system level effects<< |
| FbrakeSituation | Brakes1 | StuckNotEngaged | :Reduced_or_no_deceleration |
| FbrakeSituation | Brakes1 | StuckEngaged | :>>no system level effects<< |
| FbrakeSituation | BrakesECU1 | MissingCommand | :Reduced_or_no_deceleration |
| FbrakeSituation | BrakesECU1 | UntimelyCommand | :>>no system level effects<< |
| FbrakeSituation | RetarderECU1 | MissingCommand | :Reduced_or_no_deceleration |
| FbrakeSituation | RetarderECU1 | UntimelyCommand | :>>no system level effects<< |
| FbrakeSituation | TransmissionsECU1 | MisingClutchCommand | :Reduced_or_no_deceleration |
| FbrakeSituation | Engine1 | LowTorque | :>>no system level effects<< |
| FbrakeSituation | Engine1 | HighTorque | :>>no system level effects<< |

Figure 5.9: Hazard analysis for the driving situation "F-brake"

An evaluation of the results yields that, for the faults modeled and the considered driving situations, the tables contain no false positives and false negatives.

| Scenario | Part | Failure Mode | Hazard / Impact |
|---|---|---|---|
| DriveSituation | CrankShaft1 | Broken | :Reduced_or_no_acceleration |
| DriveSituation | Clutch1 | ClutchStuckOpened | :Reduced_or_no_acceleration |
| DriveSituation | Clutch1 | ClutchStuckClosed | :>>no system level effects<< |
| DriveSituation | GearBox1 | StuckReverse | :Unintended_deceleration |
| DriveSituation | GearBox1 | StuckNeutral | :Reduced_or_no_acceleration |
| DriveSituation | GearBox1 | StuckForward | :>>no system level effects<< |
| DriveSituation | Retarder1 | RetarderStuckNotEngaged | :>>no system level effects<< |
| DriveSituation | Retarder1 | RetarderStuckEngaged | |
| DriveSituation | Retarder1 | RetarderStuckEngaged | :Reduced_or_no_acceleration |
| DriveSituation | Retarder1 | RetarderStuckEngaged | :Unintended_deceleration |
| DriveSituation | Brakes1 | StuckNotEngaged | :>>no system level effects<< |
| DriveSituation | Brakes1 | StuckEngaged | |
| DriveSituation | Brakes1 | StuckEngaged | :Reduced_or_no_acceleration |
| DriveSituation | Brakes1 | StuckEngaged | :Unintended_deceleration |
| DriveSituation | BrakesECU1 | MissingCommand | :>>no system level effects<< |
| DriveSituation | BrakesECU1 | UntimelyCommand | |
| DriveSituation | BrakesECU1 | UntimelyCommand | :Reduced_or_no_acceleration |
| DriveSituation | BrakesECU1 | UntimelyCommand | :Unintended_deceleration |
| DriveSituation | RetarderECU1 | MissingCommand | :>>no system level effects<< |
| DriveSituation | RetarderECU1 | UntimelyCommand | |
| DriveSituation | RetarderECU1 | UntimelyCommand | :Reduced_or_no_acceleration |
| DriveSituation | RetarderECU1 | UntimelyCommand | :Unintended_deceleration |
| DriveSituation | TransmissionsECU1 | MisingClutchCommand | :Reduced_or_no_acceleration |
| DriveSituation | Engine1 | LowTorque | :Reduced_or_no_acceleration |
| DriveSituation | Engine1 | HighTorque | :Increased_acceleration |

Figure 5.10: Hazard analysis for the driving situation "Drive"

## 5.6  Case Study: Model-based Impact Analysis

The hazard analysis described above yields the consequences of faults in terms of deviations in the motion of the vehicle, more specifically, deviations of its acceleration. These deviations may be influenced by the friction and slope of the road, but are, otherwise, independent of interactions with other objects in the environment. In other words, they lie at the interface between the vehicle and its environment.

Determining the impact of this deviation on the environment requires a representation that can express the location and motion of the vehicle as well as other objects in this environment as a basis for inferring the potential of collisions. As before, this analysis is carried out for different scenarios, where scenarios in this phase are seen as different spatial configurations of the vehicle and other objects (see Figure 5.2). Besides their (potential) spatial extension, objects have an associated type (which influences the severity of the impact).

As we saw above, hazards are qualitative, and so are the different spatial configurations in the environment, which represent classes of specific real situations, such as "street with people on the sidewalk" and "approaching exit on a freeway". As a consequence, the required spatial representation has to be very abstract and qualitative, as described in the following section.

### 5.6.1  Spatial Representation

As opposed to other work that exploits spatial reasoning for exploring trajectories of moving objects and their spatial relations and predicting collisions based on particular situations (e. g., [WDFN07]), we need to represent archetypes of situations, possible ranges of motions, and the potential of collisions.

To approach this and derive a simplified representation, we first abstract from the road as a 3D object: Although it may go uphill and downhill, the 3rd dimension is eliminated and only expressed as an attribute **slope** of the road, which influences the motion of the vehicle through gravitational force, which is already covered by the vehicle model.

Secondly, we "rectify" the possibly winding trajectory: Although the road (or, more generally, the intended trajectory of the vehicle, as in "exiting from a freeway") may have curves, which influences the impact (e. g., at high speeds), we also turn this into a (Boolean) attribute of the road, indicating whether the **curvature** is significant or not, and transform the space by turning the vehicle trajectory into one coordinate axis, $\sigma$, and the orthogonal distance from the road the other coordinate, $\delta$, with the initial location of the vehicle in the origin, as illustrated by Figure 5.11.

Next, we abstract this space according to the distinctions that appeared in the natural language scenario descriptions supplied by the industrial partner, i.e. we discretize $R^2$ to a level that captures the qualitative distinctions needed to characterize locations and that is able to infer a potential collision due to the (qualitatively) deviating motion of the vehicle. As an initial solution we chose the grid depicted in Figure 5.12. The grid is defined by qualitative positions 0 (at the vehicle), **c**lose, **m**edium, **f**ar (both in front of and behind the vehicle) for $\sigma$, i. e., along the vehicle trajectory, and **s**traight, **r**ight-of, **m**edium-**r**ight-of, **f**ar-**r**ight-of (and the same for left) for $\delta$, the distance from the trajectory. The vehicle's initial position will always be in (0, s), while pedestrians may cover the r-strip, or a median be located in the l-strip. Figure 5.13 depicts how different spatial configurations are represented using this grid (note that curves are not visible in the graphical representation).

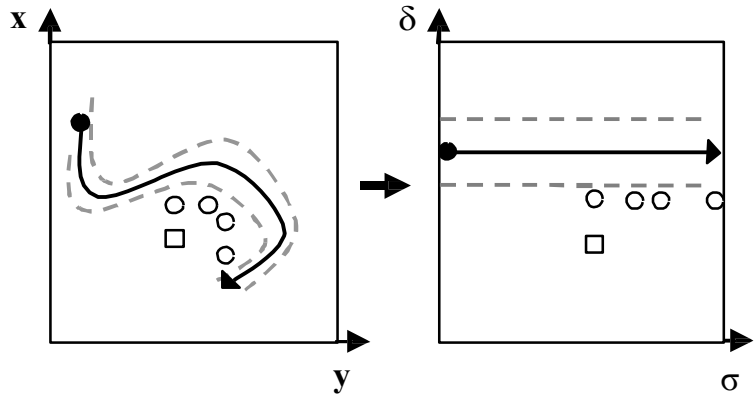

Figure 5.11: Rectifying the road

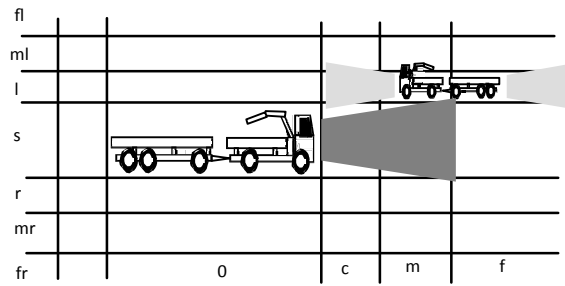

Figure 5.12: The qualitative spatial representation

### 5.6.2 Model for Impact Analysis

The concept behind the environment model is to map hazards (expressed as deviations of acceleration and, in future extensions, steering deviations) to potential regions of impact and to determine the impact in terms of **potential collisions**. Since a collision means that two objects are in the same location, potential collisions are obtained as non-empty intersection of impact ranges of the vehicle and other objects, where the impact range of the vehicle is the set of the potential locations after the initial situation. The impact range is influenced by the driving situation (direction and magnitude of speed, accelerating or decelerating), the curvature of the road, and the presence of hazards, i.e., deviations in acceleration). At this time, we ignore the influence of another factor, the mass of the vehicle, which may vary, especially for trucks.

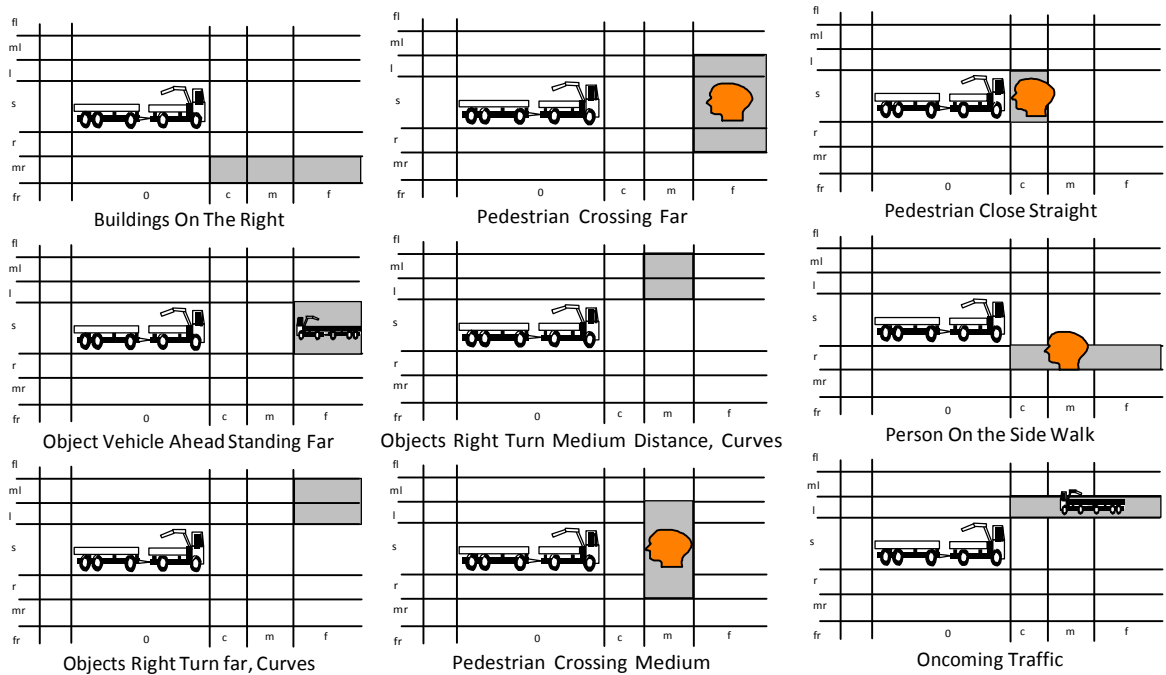

Figure 5.13: Overview of spatial configurations

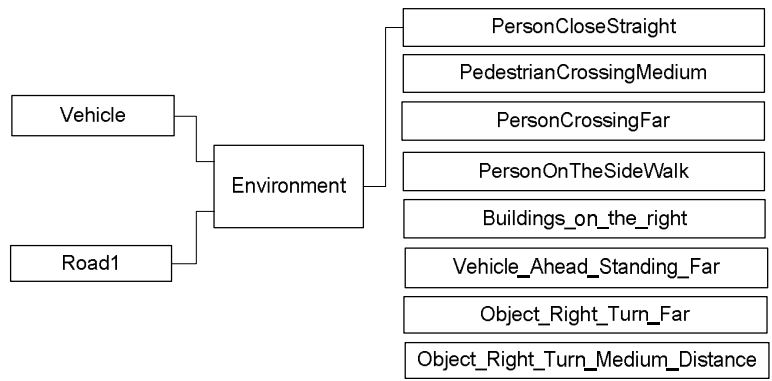

Figure 5.14: The spatial reasoning components. For each reasoning step, only one element from the right, which represents a spatial configuration, is connected to form a specific spatial configuration as part of a scenario.

As shown in Figure 5.14 (a screen shot from Raz'r), the model that produces the required inferences comprises

- the **vehicle**,

- the **road**,

- another **object** (which can represent an aggregation of individual objects, such as pedestrians or buildings), and

- the **environment** as the central "component" that performs the essential part of the collision analysis.

The **environment** model is fairly straightforward: it has only an **OK model**, which has the task of **determining the impact range of the vehicle** from

- the **speed** of the vehicle,

- its **acceleration deviation**,

- a **deviation of the steering angle** (which is not within the scope of the modeled part of the system) and

- the **curvature** of the road,

which is expressed in one impact range constraint. In our current vehicle model, deviations in steering do not occur, but the impact of a positive deviation and a high speed may lead the vehicle away from s-locations if the road has a significant curvature. The basis for this is the introduction of the distinction between low and high speed.

The impact has to be defined as the system level effect (which is the **local effect of the environment** component) in this step of the analysis. Since the relevant impact is a potential collision, which means the impact range of the vehicle and the set of potential object locations have a non-empty intersection, these collision effects simply contain a constraint

$$\text{Equal}(vehicle.impactrange, object.location), \qquad (5.14)$$

which is satisfied unless the sets of possible values are a disjoint. As a basis for the assessment of severity, especially when persons are affected, the type of the object is important, and we define an impact "injured-persons" by adding the constraint Equal (object.type, persons).

Impact analysis is then carried out again by applying the FMEA engine to a list of faults and a set of environment conditions to determine whether or not one of the collision effects are consistent. Note that the collision occurs as a definite effect if and only if the impact range and the potential object locations are the same singleton. Based on the introduced models, we have performed two variants of impact analysis, as discussed in chapter 5.4: determining

- **impacts from hazards**, using the hazards as the fault modes and specifying scenarios as environment conditions

- **impacts from fault modes**, which are component faults and analyzed under different environment conditions.

We present the results in the next two sections.

| Name | Driving Situation | Road | Spatial Configuration | Part | Failure Mode | Hazard / Impact |
|---|---|---|---|---|---|---|
| Pedestrian crossing medium, low speed | Low Speed | No curves | Pedestrian Crossing Medium | Vehicle | deltaaminus | :>>no system level effects<< |
| | | | | Vehicle | deltaaplus | :Injury_Person |
| Pedestrian crossing far, high speed | High Speed | No Curves | Pedestrian Crossing Far | Vehicle | deltaaminus | :>>no system level effects<< |
| | | | | Vehicle | deltaaplus | :Injury_Person |
| Pedestrian crossing far, low speed | Low Speed | No Curves | Pedestrian Crossing Far | Vehicle | deltaaminus | :>>no system level effects<< |
| | | | | Vehicle | deltaaplus | :>>no system level effects<< |
| Person on the side walk, high speed | High Speed | Curves | Person On The Side Walk | Vehicle | deltaaminus | :>>no system level effects<< |
| | | | | Vehicle | deltaaplus | :Injury_Person |
| Person on the side walk | Low Speed | No Curves | Person On The Side Walk | Vehicle | deltaaminus | :>>no system level effects<< |
| | | | | Vehicle | deltaaplus | :>>no system level effects<< |
| Highway in city Curves, high speed | High Speed | Curves | Buildings On The Right | Vehicle | deltaaminus | :>>no system level effects<< |
| | | | | Vehicle | deltaaplus | :collision_with_object |
| Freeway exit ahead, high speed | High Speed | Curves | Object Right Turn Far | Vehicle | deltaaminus | :>>no system level effects<< |
| | | | | Vehicle | deltaaplus | :collision_with_object |
| Roundabout medium distance no traffic, low speed | Low Speed | Curves | Object Right Turn Medium | Vehicle | deltaaminus | :>>no system level effects<< |
| | | | | Vehicle | deltaaplus | :>>no system level effects<< |
| Freeway approaching tail of traffic jam, high speed | High Speed | No Curves | Vehicle Ahead Standing Far | Vehicle | deltaaminus | :>>no system level effects<< |
| | | | | Vehicle | deltaaplus | :collision_with_standing_vehicle |

Table 5.4: Impact analysis 1, results from hazards to impacts

### 5.6.3 Impact Analysis 1: from Hazards to Impacts

This analysis is based on the model described in the previous section. The set of fault modes is determined by the vehicle component and actually quite small, namely a positive and a negative deviation of acceleration. The result of performing this analysis is shown in Table 5.4.

| Name | Driving Situation | Road | Spatial Configuration | Part | Failure Mode | Hazard / Impact |
|---|---|---|---|---|---|---|
| Freeway exit ahead, braking, high speed | Braking, High Speed | Curves | Object Right Turn Far | CrankShaft1 | Broken | :collision_with_object |
| | | | | Clutch1 | ClutchStuckOpened | :collision_with_object |
| | | | | Clutch1 | ClutchStuckClosed | :>>no system level effects<< |
| | | | | GearBox1 | StuckReverse | :collision_with_object |
| | | | | GearBox1 | StuckNeutral | :collision_with_object |
| | | | | GearBox1 | StuckForward | :>>no system level effects<< |
| | | | | Retarder1 | RetarderStuckNotEngaged | :collision_with_object |
| | | | | Retarder1 | RetarderStuckEngaged | :>>no system level effects<< |
| | | | | Brakes1 | StuckNotEngaged | :collision_with_object |
| | | | | Brakes1 | StuckEngaged | :>>no system level effects<< |
| | | | | BrakesECU1 | MissingCommand | :collision_with_object |
| | | | | BrakesECU1 | UntimelyCommand | :>>no system level effects<< |
| | | | | RetarderECU1 | MissingCommand | :collision_with_object |
| | | | | RetarderECU1 | UntimelyCommand | :>>no system level effects<< |
| | | | | TransmissionsECU1 | MisingClutchCommand | :collision_with_object |
| | | | | Engine1 | LowTorque | :>>no system level effects<< |
| | | | | Engine1 | HighTorque | :>>no system level effects<< |

Table 5.5: Results of the impact analysis for the "freeway exit ahead" environment condition

### 5.6.4 Impact Analysis 2: from Component Faults to Impacts

The vehicle models shown in Chapter 2 and Section 5.5.4 can be composed, where the vehicle and road components provide the interface. This allows directly determining the impact as a result of the vehicle component faults (the two vehicle fault modes have to be deleted for this analysis, because they represent the effect of component faults at the vehicle level). We present the result for the environment condition "Freeway exit ahead" in Table 5.5.

The most important cases for the complete impact analysis are listed below (see Figure 5.13 for a visual representation of the cases):

- Pedestrian crossing medium
- Pedestrian crossing far
- Person on the sidewalk
- Highway in city
- Freeway exit ahead
- Roundabout medium distance no traffic
- Freeway approaching tail of traffic jam

## 5.7 Limitations of the Case Study

In the analysis presented in Section 5.5 and 5.6, there is still missing a representation and inferences of an impact on persons and objects **inside the vehicle**. However, the modeling principles with its clear boundaries between the physical behavior of the vehicle and its interaction with the environment and the algorithmic solution apply to this and other kinds of impacts as well, e. g., exposure to heat or electrical charges.

Within the scope of the modeled system, the current gear box model is overly simplified, i. e., not distinguishing between different forward gears. Moreover, there are more spatial configurations to be included. The ECU models could also be refined and analyzed in the context of different ECU architectures. Furthermore, we did not yet consider impacts in the rear of the vehicle, e. g., unintended braking might cause a subsequent vehicle colliding with the one under analysis. Finally, the influence of slope and surface friction of the road on the impact is included in the model but currently not exploited by the analysis.

Including more components in the vehicle model, such as electrical ones, would extend the value of the case study. For instance, a fault in the electrical engine seen as a generator may ultimately affect the function of the steering pump and potentially create possible deviations in the steering angle. On the other hand, it may also turn into an electric motor, adding torques to the power train. A more detailed model of braking may also reveal an impact of faults that lets the vehicle yaw to one side, as described in [SF12b].

## 5.8 Summary

By taking the third perspective, we presented automated safety analysis based on behavior models of physical and software components. The approach used is based on compositional modeling, qualitative deviation models, and automated prediction of effects. We described how we built a model for the case study and illustrated results of the reasoning procedure consisting of hazard and impact analysis. The case study is a proof of concept that automation of the analysis along the lines of Figure 5.2 is feasible. Given the

- **component model library** which is not system specific but reusable for a class of systems in a respective domain, and the

- **spatial configurations** which are also reusable but related to the type of impact to be studied,

the user of the method is only required to supply

- the **structural description** which is the essential information about the system, and
- definitions of **hazards** if hazard analysis is to be performed.

These definitions will be relevant to a whole class of systems. Based on them, analysis results can be obtained automatically.

This approach was validated by creating models of the main physical components and ECUs of a truck drive train and using them to infer the potential of hazards from assumed component faults, where, due to the functions of the system, hazards are deviations from nominal acceleration (and deceleration). The models are generic and reusable in different contexts and for different system structures. In summary, the work provided a proof of concept.

Using the same algorithm, we also fully automated the analysis of the impact on the environment, starting either from the hazards or again from component faults (Section 5.1). In the context of the case study, the relevant impacts are collisions of the vehicle with persons and objects. This analysis requires a coarse-grained spatial representation and some basic inferences about the potential motion of the vehicle relative to the location of other objects. The developed model is generic, as well, and allows for an easy extension of the set of spatial configurations of objects. Of course, it is restricted to the class of impacts that result from the motion of the vehicle, which is compliant with the scope of ISO 26262.

Under the qualitative and worst-case perspective required by safety analysis, the results generated by the automatic analysis appear to be complete and sound, i. e., including neither false negatives nor false positives. Obviously, they include no results that could not have been generated manually. But this should be seen as a positive, rather than a negative feature. After all, the objective of this work is not producing insights beyond the current practice of engineers, but reducing the work load in safety analysis by automating mechanistic reasoning steps in the analysis. The production of the tables shown in this report takes less than a second, whereas producing them manually will require several person hours of work.

# 6 Discussion and Future Work

## 6.1 Relationships between the Three Perspectives

**From First to Second Perspective** The semi-formal way of how hazards are encoded in the first perspective (Section 3.3) can be coupled with the way defective transitions are modeled in the second perspective (Section 4.3). Hence, from the viewpoint of qualitative state machine modeling, the first and second perspectives are tightly related and compatible. This is also shown in Table 3.3 which relates the first and second perspective by means of the system interface, both models refer to.

**From Second to Third Perspective and Back** For an integration of the second and third perspective, the driving situations in Table 5.2 can be aligned with the outputs of the environment state machine (Figure 4.3). This state machine could be utilized to provide a more compact and yet general scenario model. The OK model matches the glass-box version of $f^*_{use}$, the fault models match the pendants of $f^*_{fail}$. The atomic transitions in the state machine based perspective (e.g., Figure 4.4b), particularly the defective transitions, i.e., parts of the failure model, can be well investigated and justified by the FMEA procedure explained in Section 5.5. Similarly, the third perspective can be utilized to derive common cause failures and represent it in the system-level state machine $f^I$. Beyond the hazard and impact analysis explained in Section 5, Section 4 provides guidance for hazard treatment at a state machine level. Steps 3 and 4 in Section 4.2 match the reasoning step from hazards to impacts explained in Section 5.6 and, particularly, Section 5.6.3.

## 6.2 Deriving Safety Requirements for Embedded Software

In the perspectives presented above, the models were used for determining hazards and their impact on the environment, i.e., for analysis only. However, regarding the perspective shown in Figure 5.1, the model also forms the basis for the derivation of safety requirements and, hence, could contribute to re-design for safety. Based on the formalism introduced in Sections 5.2–5.4, we illustrate this potential in an abstract way: First, in the analysis step, a particular physical scenario, $S_P$, (say, heavy braking on a slope) is mapped to the input channel of the software by the physical model, $R_{PS}$ (which can be $R_{OK}$ or some $R_{MA_i}$ representing a faulty component), as a set of sensor signals, or, rather, ranges of sensor signals (e.g., pressure, wheel speeds, etc.),

$$I_S = \pi_I(R_{PS} \quad S_P), \tag{6.1}$$

where $\pi_I$ denotes the projection to the input channels $I$ of the embedded software.

The software model $R_{SW}$ needs to determine the respective output in terms of actuator signals (e. g., to the valves controlling the braking)

$$O_S = \pi_O(R_{SW} \quad I_S), \qquad (6.2)$$

where $\pi_O$ is the projection to the output channel. Based on the scenario $S_P$ and $O_S$, which is the input to the physical system, $R_{P\,S}$ determines the behavior of the physical system with respect to its environment:

$$B_E = \pi_E(S_P \quad R_{P\,S} \quad O_S), \qquad (6.3)$$

where $\pi_E$ is the projection to the interface of the physical system to the environment (e. g., too high deceleration), which may then lead to a relevant impact on the environment.

On this basis, safety requirements for the embedded software may be determined by "back-propagating" a safety requirement on the behavior of the physical system to the software: avoiding the impact by avoiding the hazard $B_E$ establishes a revised system response $B_E^t$, (e.g., the complement of $B_E$ or a subset of it). $R_{P\,S}$ infers a required modified software output in scenario $S_P$

$$O_S^t = \pi_O(B_E^t \quad R_{P\,S} \quad S_P), \qquad (6.4)$$

i. e., the requirement on the modified software model $R_{SW}^t$

$$\pi_O(R_{SW}^t \quad I_S) \subseteq \pi_O(B_E^t \quad R_{P\,S} \quad S_P), \qquad (6.5)$$

or describing the software as a function $F_{SW}^t : I \to O$ with

$$F_{SW}^t : \pi_I(R_{PS} \quad S_P) \to \pi_O(B_E^t \quad R_{P\,S} \quad S_P) \qquad (6.6)$$

satisfies the requirement.

The state machine approach used in the second perspective (Section 4.3) always refers to a specific system boundary. Now, this approach can be applied to software as follows: In Section 4.3, we exemplified the state machine based on the boundary of the physical system (Figure 5.1). Having the derived function $F_{SW}^t$, we can enter the procedure described for the second perspective, starting with the derivation of a nominal state machine, etc. However, still a point of future work is how to perform Steps 3 and 4 in detail solely for software components using traditional software FMEA or FTA approaches (Section 1.2).

# 6 Discussion and Future Work

We presented three perspectives of safety analysis substantially based on the results of the project *Efficient Hazard and Impact Analysis for Automotive Mechatronics Systems* on behalf of ITK Engineering AG, Stuttgart-Vaihingen. On the basis of the characteristics of these three perspectives as summarized in Sections 3.4, 4.4 and 5.8, we conclude our report as follows:

- The first perspective leverages semi-formal specification of *hazards* and *safety goals* using reusable textual patterns and allows optimizing hazard tables and their content by identifying logical relationships (e.g., refinement between hazards) among and between these two classes of assertions.

- The second perspective leverages state machines for hazard analysis and aims to ease the management of large hazard tables as well as to allow for additional analyses and the conduct of subsequent safety engineering tasks (e.g., the constructive treatment of hazards). This behavior-oriented perspective is suited to be applied early in the safety life cycle.

- The third perspective automatically performs integrated hazard and impact analysis using a model of the system structure and qualitative models of physical and software component behavior. It also considers the system environment using the same modeling paradigm and applies an automated FMEA algorithm for end-to-end cause-effect reasoning.

The first two perspectives focused holistic aspects of systems, control and software engineering and leveraged the idea that speaking about behavioral properties of the whole system, one can postpone to distinguish physical signals and components from the software counterparts to later development phases. We consider this as an advantage for early stage hazard and impact analysis and for reusing such system models independent of technical solutions. Nevertheless, the consideration of the system structure or physical architecture design including modularization into mechanical, electrical and software parts as focused by the third perspective is mandatory to capture the fault space and to justify the more abstract reasoning results. Overall, qualitative physical modeling plays an important role in all of the three perspectives, even if formalized in a different fashion. Section 6.1 provides some more details on the relationships of the three perspectives.

In a first step and by means of a case study, we approached the criteria for an appropriate hazard and impact analysis methodology as characterized by the project goals and criteria in Section 1.3. As a point of future work, Section 6.2 formally sketches how to derive interface requirements for control software from the requirements for the overall physical system.

# 7 Conclusion

# Glossary

**Automotive Safety Integrity Level (ASIL)** after ISO Std. 26262, a set of safety-specific requirements for an item, e. g., a system, a component, a software module or a process activity, depending on the hazard assessment result. 14

**behavioral property** a property of a system observable at the system interface, i.e., its inputs/stimuli or outputs/reactions. 15, 65, *used by* requirement

**common cause failure** (CCF) several system failures caused by the same single or set of faults. 59

**component** role of an entity being part of a system, can itself be considered as a system. 35, 36, 63–65

**constraint** (i) a mathematical relation fulfilling some property or predicate, or (ii) just the property or predicate itself. 26, 65

**cyber-physical system** view of a system with a focus on the relationship between software components, computer networks and physical components to ultimately provide functionality to a user. 5, 15, 33, 34, *in this report also synonym for* system

**defect model** a system model capturing structural or behavioral deviations or discrepancies, e.g., failures or faults, from the specified ordinary system. 25, 64

**driving situation** gives information about the driving mode and velocity. 12, 35, 38, 44, 46, 59, 63, 67, *part of the* environment condition

**ECU** Electronic or Embedded Control Unit. 11, 34, 42, 57

**effect** a result of the performance of a transition or action, this result can be negative or unwanted, e.g., a failure or hazard, as in FMEA. 33, 45, 64, 65, *see* hazard

**environment** includes the vehicle of our case study including surrounding physical objects which can have an impact on the vehicle and vice versa. 25, 33, 35, 63–65, *generalized by* operational context

**environment condition** contains information about the driving situation, road condition, and spatial configuration. 12, 35, 36, 40, 53, 65, 67, *captured by* scenario

**environment model** compact representation of the environment consisting of many scenarios. 44, *see* scenario

**event** describes a *point in time* in which the system or its environment happens to observe or do (i. e., change) something. 16, *see* state **failure** behavioral violation of specified system functionality. 25, 26, 63–65

**failure mode** . *see* fault mode

**failure model** the defect model for the first (Section 3) and second perspective (Section 4). 59, *see* defect model

**fault** potential root cause of a failure; also called weakness, error, bug or mistake. 5, 26, 33, 43, 63

**fault mode** class of individual faults in the same location or component of the model used in the third perspective. 39, 40, *see* fault

**fault model** the defect model for the third perspective (Section 5). 38, 59, *see* defect model

**faulty behavior** . 39, 41, *see* failure

**FMEA** failure mode and effects analysis; also called *failure mode, effects and criticality analysis (FMECA)*, if the assessment of criticality in terms of type and severity of impacts is tightly related with the effects. 7, 63

**hazard** a combination of the current state and mode of operation of both, the environment and the system, e.g., a system failure or maloperation, *directly or indirectly enabling* a safety risk and, thus, potentially its occurrence in terms of a mishap. 6, 9, 15, 19, 24–26, 33, 34, 45, 63–65, *see* risk

**hazard and impact analysis** equivalent to the activity of *hazard analysis and risk assessment (HARA)* in ISO 26262 part 3. 8, 61

**impact** . 6, 12, 25, 28, 33, 35, 50, 51, 64, 65, *in this report also synonym for* mishap

**local situation** one or more states of the environment and the system, e.g., initial states. 25, 26, 29

**machine boundary** . *a synonym for* system boundary

**mishap** potential consequence or effect of a hazard; synonymous to impact, harming event, accident, incident. 25, 64, 65, *specializes* risk

**mode** (of operation) represents a specific behavior of a system *within some time period of time* and, hence, usually represents a set of states. 16, 37, 64, *see* state

**OK model** a model specifying the structure and the ordinary qualitative dynamics of a system. 59, *see* defect model

# Glossary

**operation** mathematical relation. 37, *see* constraint

**operational context** a specific part of the world capturing the environment of the system under consideration. 8, *specialized by* environment **requirement** a structural or behavioral property required to be fulfilled by the system to be built and operated or a component thereof; usually the opposite of failure or hazard. 9, 15, 63, *specialized by* safety goal

**risk** or **safety risk**: an effect, e.g., an impact or mishap, occurring at a probability of < 1 given a known scenario or state. 25, 34, 36, 64, *see* mishap

**road condition** gives information about the slope, friction, and curves. 35, 38, 44, 63, *part of the* environment condition

**safety analysis** . 33, 65, *see* hazard and impact analysis

**safety goal** a structural or behavioral property required to be fulfilled by the system to be built and operated in order to prevent from the occurrence of hazards or their impacts. 19, 24, 27, *generalized by* requirement

**safety property** . 23, 30, 67, *in this report also synonym for* safety goal

**scenario** a constraint capturing relevant initial states of the environment, e.g., environment conditions, system states. 36, 39, 50, 63, 65

**spatial configuration** gives information about location of objects in the environment. 35, 50, 52, 63, 66, *part of the* environment condition

**specification** an informal (natural language, textual), semi-formal (controlled textual patterns) or formal (mathematically defined) artifact, e. g., a document, containing a description or model of the considered system. 15, 25, *also contains elements of type* requirement

**state** represents a valuation of a set of attributes of a system or component *within some time period of time*. 16, 64, 65, *see* event

**system** a technical entity (i) consisting of interconnected components, (ii) exhibiting physical behavior and delivering a specified functionality, (iii) the subject of safety analysis. 15, 25, 33, 63–65

**system boundary** the boundary between the system and its environment to define the system interface. 19, 22, 26, 60, 65, *see* system interface

**system interface** a set of shared phenomena such as, e. g., communication channels, ports, points of interaction, as a syntactic basis or a vocabulary to specify the behavior of a system. 15, 22, 24, 25, 63, 65, *see* system boundary

**terminal** element of a system model which is needed to propagate effects between two other components, particularly used to propagate effects across the system boundary. 37, *see* system boundary

# List of Figures

# List of Tables